\documentclass[twocolumn,english,showpacs,preprintnumbers,amsmath,amssymb]{revtex4}
\usepackage{graphicx}

\makeatletter

\usepackage{dcolumn}
\usepackage{bm}

\usepackage{graphics}
\usepackage{babel}
\makeatother
\begin{document}

\title{Microscopic Theory of Superconductor-Constriction-Superconductor Josephson Junctions in a Magnetic Field}

\author{A. Gumann}
\author{T. Dahm$^\dagger$}
\author{N. Schopohl}

\affiliation{Institut f\"ur Theoretische Physik, Universit\"at T\"ubingen\\
 Auf der Morgenstelle 14, D-72076 T\"ubingen, Germany}

\date{July 25, 2007}

\begin{abstract}
Self-consistent solutions of microscopic Eilenberger theory are presented for a two-dimensional model of a superconducting channel with a geometric constriction. Magnetic fields, external ones as well as those caused by the supercurrents, are included and the relevant equations are solved numerically without further assumptions. Results concerning the influence of temperature, geometric parameters, of $\kappa=\lambda_L/\xi_0$ and of external magnetic fields on the Andreev bound states in the weak link and on the current-phase relation are presented. We find that the Andreev bound states within the junction obtain peculiar substructure when a finite supercurrent flows. As long as the London penetration depth is comparable to or bigger than the extension of the constriction, the Josephson effect is independent of $\kappa$. Furthermore, the weak link is very insensitive to external magnetic fields. Features restricted to a self-consistent calculation are discussed.
\end{abstract}

\pacs{74.25.Jb, 74.45.+c, 74.50.+r}

\maketitle

\section{\label{sec:intro}Introduction}

Since their discovery, Josephson effects play a central role in the investigation of the superconducting state because of their fundamental as well as technical relevance. Josephson effects occur whenever two extended superconducting banks are weakly coupled and can be observed, among many other systems, across a geometric constriction of a superconductor with lateral dimensions much smaller than the coherence length. This kind of Josephson junction, usually referred to as ScS junction, has promising properties for highly sensitive magnetometers~\cite{Shn01}. Recent progress in microfabrication technique allows to pattern high quality niobium thin films with structures small enough for constriction-type Josephson junctions~\cite{Lam01,Bou01}. At the same time, a ScS Josephson junction is a useful basis for theoretical considerations because only the properties of the superconductor and its geometry are relevant.

Superconducting point contacts serve as a simplified model for ScS junctions and have been investigated by many authors. An overview of many useful results for the ac and dc Josephson effect can be found in the reviews by Likharev~\cite{Lik01} and Golubov et al.~\cite{Gol01}. The most general result was derived by Zaitsev~\cite{Zai01} and includes a large number of conduction channels as well as asymmetry and non-BCS structure of the electrodes.
\\
Microscopic theory for the dc Josephson effect in a model mesoscopic constriction has been presented in~\cite{Mar01, Lev01} with the restriction to a small number of conducting channels in the constriction. This model clarified the relevance of Andreev bound states for the supercurrents in the junction. Analytic solutions of microscopic theory for the dc Josephson effect in a ballistic superconducting microconstriction have been presented in~\cite{Zar01}. In this approach, the influence of the constriction length on the critical current has been calculated self-consistently for temperatures close to $T_c$.

As already stated by Kulik~\cite{Kul01} and Ishii~\cite{Ish01,Ish02}, a more detailed physical picture of Cooper pair tunneling in Josephson junctions involves Andreev reflection~\cite{And01,And02} at the interfaces between the weak link and the electrodes. Due to this picture, Andreev bound states in Josephson junctions represent electron and hole waves that bounce back and forth between the two electrodes constructively, transferring one Cooper pair in every cycle. Thus, Cooper pair transfer across Josephson junctions is mediated by Andreev bound states. With this view in mind, detailed knowledge of the local density of states (LDOS) in the vicinity of the weak link is essential. This fact has led to the consideration of the LDOS in different types of Josephson junctions: among others, see \cite{Mar01,Lev01} for ScS, \cite{Kul01,Ish01,Ish02} for SNS, \cite{Fur00,Fur01,Fur02,Mil01} for SNS and SIS and \cite{Zik01} for SNS and SFS with s- and d-wave pairing symmetry (N = normal conductor, I = isolator, F = ferromagnet).

In the present paper, we examine a two-dimensional model of a ScS Josephson junction in order to investigate the dependence of the dc Josephson effect on temperature, on geometric parameters as well as the influence of magnetic fields, external ones as well as those generated by the supercurrents flowing in the junction. Therefore, we will numerically solve microscopic Eilenberger theory self-consistently for a conventional (s-wave) superconducting junction in the clean limit. In order to achieve electromagnetic self-consistency as well, we employ a convenient method to calculate the magnetic field evoked by the supercurrents. This will allow us to derive transport properties for a realistic geometry at arbitrary temperature and arbitrary external magnetic field. Furthermore, the method used to solve microscopic Eilenberger theory self-consistently including magnetic fields presented in this paper can be applied to more complex geometries and thus might be of more general interest.

The paper is organized as follows. In the next section, we introduce our geometric model. In section~\ref{sec:theory}, we present the theoretical background for our calculations and explain how we solve the relevant equations. In section~\ref{sec:comp}, we compare our results to previous ones in order to test our methods. In section~\ref{sec:jeffectinscs}, we present representative results and explain how the Josephson effect emerges in ScS junctions. In the following, we consider the influence of the temperature and geometric parameters (section~\ref{sec:geometry}), of $\kappa=\lambda_L/\xi_0$ (\ref{sec:kappa}) and of external magnetic fields (\ref{sec:bext}). Finally, in section~\ref{sec:concl}, we close with some concluding remarks.

\section{\label{sec:model}Model Geometry}

\begin{figure}
\begin{center}
\includegraphics[width=0.95\columnwidth, keepaspectratio]{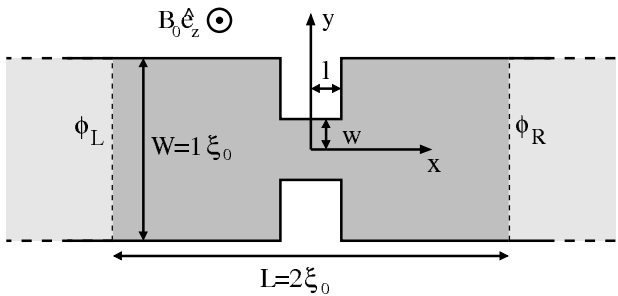}
\end{center}
\caption{\label{cap:geometry} Two-dimensional model for a ScS Josephson junction consisting of a long channel (width $W=1\xi_0$ throughout the paper) with a constriction. The constriction is defined by the parameters $l$ and $w$. Our calculations have been carried out for a section of the channel with length $L=2\xi_0$ (dark grey area) and fixed phase difference of the order parameter $\Delta\phi=\phi_R - \phi_L$. The magnetic field is oriented perpendicular to the image plane (along the $z$ axis).}
\end{figure}

For the calculations presented in this paper, we use a two-dimensional model of a ScS Josephson junction as depicted in Fig.\,\ref{cap:geometry}. The geometry consists of a long superconducting channel which is narrowed down by a constriction. We will calculate all relevant quantities on a section of the channel enclosing the constriction with length $L=2\xi_0$  ($\xi_0=\hbar v_F/\Delta_\infty(T=0)$ is the coherence length). On the left and right end of this section, the phase of the order parameter will be set to $\phi_{L,R}$. The width of the channel will be set to $W=1\xi_0$ throughout the paper whilst the constriction is defined by the parameters $l$ and $w$ as can be seen in the figure. The external magnetic field $\vec{B}_0=B_0\,\hat{e}_z$ is oriented perpendicular to the two-dimensional structure.

Since we assume the Fermi surface to be cylindrical with the symmetry axis parallel to the $z$ direction, we can restrict our calculations to the $xy$ plane. The results are applicable to films of arbitrary thickness with an appropriate value for the London penetration depth~\cite{Pea01}. Our results correspond to the symmetry plane at half the extension of the sample in the $z$ direction and, accordingly, magnetic fields are oriented parallel to $\hat{e}_z$.

\section{\label{sec:theory}Quasiclassical Eilenberger Theory}

\subsection{\label{sub:gapcurr}Self-Consistency Equations}

Our calculations are based on quasiclassical Eilenberger theory~\cite{Eil01} which we solve self-consistently including the magnetic vector potential. To obtain Josephson weak link behaviour, it is essential to calculate the order parameter self-consistently. A self-consistent solution contains local modulations of the order parameter amplitude and can thus yield local suppression of superconductivity. Local suppression of superconductivity in turn constitutes a Josephson weak link.
\\
To solve quasiclassical Eilenberger theory self-consistently, we have to find solutions for the order parameter $\Delta(\vec{r})$ and the supercurrent $\vec{\jmath}_s(\vec{r})$ which satisfy the two self-consistency conditions:
\begin{eqnarray}
\label{eq:gapeq}
\Delta(\vec{r})
&=&
2\pi V N(0) k_B T\sum_{0<\varepsilon_n<\omega_c}\big<f(\vec{r},\vec{k}_F,i\varepsilon_n)\big>_{FS}
\\
\label{eq:curreq}
\vec{\jmath}_s(\vec{r})
&=&
4\pi e N(0) k_B T\sum_{0<\varepsilon_n<\omega_c}\big<\vec{v}_F\cdot g(\vec{r},\vec{k}_F,i\varepsilon_n)\big>_{FS}\quad
\end{eqnarray}
These two equations, the gap equation and the current equation, contain the components of the quasiclassical propagator $f(\vec{r}(x),\vec{k}_F,i\varepsilon_n)$ and $g(\vec{r}(x),\vec{k}_F,i\varepsilon_n)$ which, in turn, can be expressed in terms of two complex quantities $a(x)$ and $b(x)$ called the Riccati amplitudes~\cite{Schop01,Schop02}:
\begin{eqnarray}
\label{eq:riccatiparamf}
f(\vec{r}(x),\vec{k}_F,i\varepsilon_n)
&=&
\frac{2\,a(x)}{1+a(x)b(x)}
\\
\label{eq:riccatiparamg}
g(\vec{r}(x),\vec{k}_F,i\varepsilon_n)
&=&
-i\,\frac{1-a(x)b(x)}{1+a(x)b(x)}
\end{eqnarray}
The Riccati amplitudes $a(x)$ and $b(x)$ are solutions to the Riccati differential equations:
\begin{subequations}\label{eq:riccatipde}
\begin{eqnarray}
\hbar v_{F}\partial_{x}\, a(x)+[2\tilde{\varepsilon}_{n}+\Delta^{\dagger}(\vec{r}(x))\, a(x)]\, a(x)-\Delta(\vec{r}(x)) & = & 0\nonumber
\\
\label{eq:riccatipdea}
\\
\hbar v_{F}\partial_{x}\, b(x)-[2\tilde{\varepsilon}_{n}+\Delta(\vec{r}(x))\, b(x)]\, b(x)+\Delta^{\dagger}(\vec{r}(x)) & = & 0\nonumber
\\
\label{eq:riccatipdeb}
\end{eqnarray}
\end{subequations}
The Riccati differential equations have to be solved
along real space trajectories $\vec{r}(x)$ pointing in the direction
of the Fermi wave vector $\vec{v}_{F}(\vec{k}_F)$ using the modified Matsubara frequencies $\tilde{\varepsilon}_{n}$:
\begin{eqnarray}
i\tilde{\varepsilon}_{n}
&=&
i\varepsilon_{n}+\frac{e}{c}\,\vec{v}_{F}\cdot\vec{A}(x)
\\
\varepsilon_{n}
&=&
(2n+1)\,\pi\, k_B T
\end{eqnarray}
The function $\Delta(\vec{r}(x))=|\Delta(\vec{r}(x))|\,e^{i\phi(\vec{r}(x))}$ in the Riccati equations is the complex order parameter. The gradient of the order parameter phase $\vec{\nabla}\phi(\vec{r})$ together with the magnetic vector potential $\vec{A}(\vec{r})$ consitute the gauge-invariant superfluid velocity $\vec{v}_s(\vec{r})$:
\begin{eqnarray}
\label{eq:Dopplervs}
\vec{v}_s(\vec{r})
&=&
\frac{1}{2m}\left(\hbar\vec{\nabla}\phi(\vec{r})-\frac{2e}{c}\vec{A}(\vec{r})\right)
\end{eqnarray}
In the following subsection, we will describe the procedure how we calculate the magnetic vector potential $\vec{A}(\vec{r})$. The distribution of the order parameter phase $\phi(\vec{r})$ follows from the solution of the gap equation which, together with the current equation~(\ref{eq:curreq}), minimizes free energy. Thus, the way we calculate the magnetic vector potential makes use of the freedom of gauge, but after choosing a certain way to calculate $\vec{A}(\vec{r})$, the order parameter phase is uniquely defined by the self-consistent solution of the gap and the current equation.

To obtain stable solutions, the Riccati equation for $a(x)$ has to be integrated along $\vec{v}_{F}$ whereas the Riccati equation for $b(x)$ in the opposite direction. Appropriate initial values for the integration of the Riccati differential equations can be constructed by dropping the gradient terms. For $\varepsilon_{n}>0$, this yields
\begin{subequations}
\label{eq:abasymp}\begin{eqnarray}
a(-\infty) & = & \frac{\Delta(-\infty)}{\tilde{\varepsilon}_{n}+\sqrt{\tilde{\varepsilon}_{n}^{2}+|\Delta(-\infty)|^{2}}}\label{eq:aasymp}
\\
b(+\infty) & = & \frac{\Delta^{\dagger}(+\infty)}{\tilde{\varepsilon}_{n}+\sqrt{\tilde{\varepsilon}_{n}^{2}+|\Delta(+\infty)|^{2}}}\label{eq:basymp}
\end{eqnarray}
\end{subequations}
For the evaluation of the right hand side of the gap and the current equation, the components of the quasiclassical propagator $f(\vec{r}(x),\vec{k}_{F},i\varepsilon_{n})$ and $g(\vec{r}(x),\vec{k}_F,i\varepsilon_n)$ constructed with the Riccati amplitudes according to equations (\ref{eq:riccatiparamf}) and (\ref{eq:riccatiparamg}) have to be averaged over the Fermi surface of the superconducting material (denoted with $\langle\cdots\rangle_{FS}$). As stated above, we assume the Fermi surface to be cylindrical with the symmetry axis along the $z$ direction. Finally, these averaged quantities have to be summed over all Matsubara frequencies $\varepsilon_n$ smaller than the cutoff $\omega_c$. For our calculations, we use a cutoff of $\omega_c = 50\,k_B T_c$ which is sufficiently large for the results to be independent of $\omega_c$.

Apart from the left and right end of the channel section, all other boundaries of the geometry are assumed to be impenetrable. Thus, we can apply the general boundary conditions for the Riccati amplitudes given by Shelankov and Ozana~\cite{She01} in a very simple form. Vanishing transmissibility leads to specular reflection of the trajectories (outgoing angle is equal to incident angle) and to the conservation of the Riccati amplitudes.
\\
The optical behaviour of the trajectories makes it necessary to implement a two-dimensional raytracing procedure. Depending on position and angle, this leads to multiple reflections of the trajectories within the geometry. A similar raytracing procedure for the Riccati equations has been used in~\cite{Ini01} to calculate the local density of states at polygonal boundaries of d-wave superconductors. In our case, the trajectories can be terminated either if the left or right channel end have been reached or if the length of the trajectory is sufficient for the results of the integration to be independent of the initial values.

To solve the above equations, we discretize the area of the channel section and solve the equations at the grid nodes. Numerical costs can be reduced significantly by a local variation of the grid width. In the direct vicinity of the constriction, a rather fine grid is needed whereas less nodes are sufficient in the channel.

\subsection{\label{sub:mag}Magnetic Fields}

To account for magnetic fields, the two self-consistency equations (\ref{eq:gapeq}) and (\ref{eq:curreq}) have to be completed by the Maxwell equation
\begin{eqnarray}
\label{eq:maxwell}
\vec{\nabla}\times\vec{B}=\frac{4\pi}{c}\,\vec{\jmath}_s
\end{eqnarray}
and the definition of the magnetic vector potential
\begin{eqnarray}
\label{eq:magvecpotdef}
\vec{\nabla}\times\vec{A}=\vec{B}
\end{eqnarray}
The total magnetic vector potential has two contributions. First, there is the contribution $\vec{A}_0$ which represents the externally applied magnetic field. Second, the supercurrents flowing in the junction cause a magnetic vector potential $\vec{A}_c$. Combining both yields the total magnetic vector potential:
\begin{eqnarray}
\label{eq:Atot}
\vec{A}=\vec{A}_0+\vec{A}_c
\end{eqnarray}
Choosing Coulomb gauge $(\vec{\nabla}\vec{A}=0)$, it follows from~(\ref{eq:maxwell}) and~(\ref{eq:magvecpotdef}) for the contribution evoked by the supercurrents that
\begin{eqnarray}
\label{eq:Apoisson}
-\Delta\vec{A}_c=\frac{4\pi}{c}\vec{\jmath}_s
\end{eqnarray}
In order to solve this Poisson equation for the magnetic vector potential $\vec{A}_c$, we use the two-dimensional Green's function of the Laplacian. This leads to
\begin{eqnarray}
\label{eq:Afromcurrs}
\vec{A}_c(\vec{r})
&=&
-\frac{4\pi}{c}\int\frac{1}{2\pi}\ln{\left|\vec{r}-\vec{r}'\right|}\cdot\vec{\jmath}_s(\vec{r}')\cdot d^2 r'
\end{eqnarray}
Free space boundary conditions for $r\rightarrow\infty$ are contained in the Green's function whereas the behaviour in the vicinity of the sample is determined by the self-consistency conditions (\ref{eq:gapeq}) and (\ref{eq:curreq}), see subsection~\ref{sub:iterations}. In a two-dimensional discretization of the geometry, this integral can easily be solved analytically for each grid cell and then summed over all cells.

In order to have consistent behaviour of the magnetic vector potential at the connection to the long channel, we assume the currents to be constant away from the constriction (for $|x|\geq L$) and also integrate over these in equation~(\ref{eq:Afromcurrs}).
\\
After calculating the magnetic vector potential $\vec{A}_c(\vec{r})$ from the supercurrents via eq.\,(\ref{eq:Afromcurrs}), we make use of the freedom of gauge. The phase gradients of the order parameter $\vec{\nabla}\phi(\vec{r})$ together with the total magnetic vector potential $\vec{A}(\vec{r})$ lead to the final form of the supercurrents (see eq.\,(\ref{eq:Dopplervs})). This implies that the resulting phase gradients depend on the form of the magnetic vector potential and vice versa. To minimize both the phase gradients and the magnetic vector potential, we shift the vector potential $\vec{A}_c(\vec{r})$ to minimal absolute values. This can be done via a simple homogeneous offset which neither changes the corresponding magnetic field $\vec{B}=\vec{\nabla}\times\vec{A}$ nor conflicts with the Poisson equation\,(\ref{eq:Apoisson}). Minimizing the absolute values of the phase gradient and the magnetic vector potential leads to more robust convergence of the iterative procedure which we use to solve the gap and the current equation. This iterative procedure will be described in detail in subsection~\ref{sub:iterations}.

To complete the total magnetic vector potential, we include external magnetic fields $\vec{B}_0=B_0\,\hat{e}_z$ via $\vec{A}_0=-B_0 y\,\hat{e}_x$ which complies with the Coulomb gauge chosen for the magnetic vector potential $\vec{A}$ and which minimizes absolute values.

\subsection{\label{sub:phaseboundary}Boundary Conditions for the Order Parameter}

In the problem considered in the present work, two factors leading to pair breaking are relevant. First, a phase difference is applied which leads to a transport current and thus suppresses superconductivity. Second, the external magnetic field leads to screening currents and pair breaking. In order to apply correct boundary conditions at the connection to the channel, these two effects have to be clearly separated.

Since we study Josephson junction behaviour, the maximum transport current will be given by the Josephson critical current which is much lower than the critical current of a massive superconducting lead. Thus, the transport current will not significantly lower the gap amplitude in the channel, far away from the constriction. For vanishing external magnetic field, it is therefore a very good approximation to use the temperature dependent bulk gap amplitude $\Delta_\infty(T)$ as boundary condition at the left and right end of the section of the channel enclosing the constriction. Furthermore, for vanishing external magnetic field, the phase is homogeneous over the cross section of the channel and a phase difference can easily be applied by setting the phase boundary conditions to $\phi_{R,L} = \pm\Delta\phi/2$.

The situation changes in the presence of an external magnetic field. The magnetic field leads to screening currents and lowers the gap amplitude in the channel. To gain correct behaviour for higher magnetic fields, it is thus essential to include the lowering of the gap amplitude in the channel. With an external magnetic field, we thus proceed in two steps. First, we self-consistently calculate the amplitude and phase of the order parameter for zero transport current (corresponding to zero phase difference) without any boundary conditions at the left and right section ends. This will give us the correct self-consistent solution with screening but without transport currents and allows for a lowering of the gap amplitude as well as for variations of the phase over the cross section of the channel. In the second step, we use the amplitude and phase of the order parameter found at the left and right section end in the first step as boundary conditions for finite transport current. Therefore, the phase has to be shifted by $-\Delta\phi/2$ at the left section end and by $+\Delta\phi/2$ at the right section end whereas the amplitude has to be retained. Herewith, the pair-breaking influence of the external magnetic field and the screening currents is included for all amplitudes of the transport current. This two-step procedure has to be repeated for every value of the magnetic field.

The gradients of the order parameter phase $\vec{\nabla}\phi(\vec{r})$ together with the magnetic vector potential $\vec{A}(\vec{r})$ constitute the gauge-invariant superfluid velocity $\vec{v}_s(\vec{r})$ given in eq.\,(\ref{eq:Dopplervs}). The current-phase relation thus has to be expressed in terms of the gauge-invariant phase difference $\gamma$ with
\begin{eqnarray}
\gamma
&=&
\Delta\phi - \frac{2\pi}{\Phi_0}\int_L^R \vec{A}\,d\vec{l}
\end{eqnarray}
where $\Phi_0=\frac{h c}{2 e}$ is the flux quantum and $L,R$ indicate the left and right section end.

\subsection{\label{sub:iterations}Iterative Procedure}

Let us now comment on how we gain the self-consistent solutions for the order parameter $\Delta(\vec{r})$, the supercurrents $\vec{\jmath}_s(\vec{r})$ and the magnetic vector potential $\vec{A}(\vec{r})$. The self-consistency equations (\ref{eq:gapeq}) and (\ref{eq:curreq}) serve as iteration rules with which we improve an initial guess on a discrete grid. In each iterative step, we first have to solve the Riccati equations (\ref{eq:riccatipde}) for all Matsubara frequencies $0<\varepsilon_n<\omega_c$ and all Fermi velocities $\vec{v}_F$ at every grid node. Second, we construct $f(\vec{r}(x),\vec{k}_F,i\varepsilon_n)$ and $g(\vec{r}(x),\vec{k}_F,i\varepsilon_n)$ from the solutions for $a(x)$ and $b(x)$ via (\ref{eq:riccatiparamf}) and (\ref{eq:riccatiparamg}) and plug these into the gap (\ref{eq:gapeq}) and current equation (\ref{eq:curreq}). Third, we calculate the magnetic vector potential $\vec{A}_c(\vec{r})$ caused by the supercurrents via (\ref{eq:Afromcurrs}) and construct the total magnetic vector potential $\vec{A}$ according to~(\ref{eq:Atot}). With the new configurations for $\Delta(\vec{r})$, $\vec{\jmath}_s(\vec{r})$ and $\vec{A}(\vec{r})$, we can attempt the next iterative cycle.

This iterative procedure has to be repeated until convergence. Two criteria can be utilized to determine the level of self-consistency of the solution. It is straightforward to observe the change of the functions $\Delta(\vec{r})$, $\vec{\jmath}_s(\vec{r})$ and $\vec{A}(\vec{r})$ in every iterative step and to continue until it is sufficiently small. Since the rate of convergence in a single iteration can be very small, a more significant approach is to verify current conservation which provides a stringent test of self-consistency~\cite{Fur03,Sol01}.

In the formulation of the microscopic Eilenberger equations used in the present work, the order parameter in the gap equation retains a complex quantity. Thus, iterating the gap equation yields both the amplitude and the phase distribution of the order parameter self-consistently. The magnetic vector potential is calculated from the currents and the external magnetic field in every iteration and the system of equations is thus solved entirely self-consistent including the magnetic vector potential.

At first sight, the free space Green's function~(\ref{eq:Afromcurrs}) for the magnetic vector potential does not provide boundary conditions for the surface of the geometry. Since the Green's function~(\ref{eq:Afromcurrs}) is embedded in a self-consistent calculation, the compliance with the boundary conditions is ensured by the gap~(\ref{eq:gapeq}) and the current equation~(\ref{eq:curreq}). A self-consistent solution of the Eilenberger equations automatically leads to $\vec{\nabla} \vec{\jmath}_s=0$ and respect of the boundary conditions at the surface of the superconductor ($\hat{n}\cdot \vec{\jmath}_s=0$ with the surface normal $\hat{n}$). Since, in the self-consistent calculation, the magnetic vector potential depends on the currents and vice versa, the correct behaviour of the magnetic vector potential including $\vec{\nabla} \vec{A}=0$ follows. Embedding the Green's function~(\ref{eq:Afromcurrs}) in the self-consistent calculation thus liberates us from the explicit formulation of boundary conditions for the magnetic vector potential.

It should be noted that an iterative algorithm usually produces stable solutions. As discussed in~\cite{Lev01}, unstable solutions corresponding to saddle points of the free energy can only be reached with more elaborate methods. However, since we are only interested in stable solutions, an iterative approach is controllable and straightforward.

\subsection{\label{sub:ldos}Local Density of States}

Once a self-consistent configuration of the order parameter, the supercurrents and the magnetic vector potential has been found, the normalized local density of states follows via
\begin{eqnarray}
\label{eq:ldos}
\frac{N(E,\vec{r})}{N(0)}
&=&
-\big<\mbox{Im}\left[ g(\vec{r},\vec{k}_F,i\varepsilon_n \rightarrow E+i\delta)\right] \big>_{FS}
\end{eqnarray}
with the definitions of $g$ and $\langle\dots\rangle_{FS}$ introduced above~\cite{foot1}.

\subsection{\label{sub:units}Normalization}

All quantities with the dimension of an energy will be normalized to $k_B T_c$. The characteristic lengthscale we use to normalize all lengths is $\xi_0=\frac{\hbar v_F}{\Delta_\infty(T=0)}$. Total currents (integrated over the cross-section of the channel) will be given in terms of $\hat{\jmath}_{tot}=\frac{j_{tot}}{e N(0) v_F k_B T_c \xi_0}$ with $v_F=|\vec{v}_F|$.
\\
With $\kappa=\lambda_L/\xi_0$ and the London penetration depth $\lambda_L^{-2}=4\pi (e/c)^2 N(0) v_F^2$, we use $\hat{\vec{B}}=\frac{\vec{B}}{\Phi_0/(1.764\pi\,\xi_0^2)}$ and $\hat{\vec{A}}=\frac{\vec{A}}{\Phi_0/(1.764\pi\,\xi_0)}$.
\\
Applying these normalizations, equation~(\ref{eq:Apoisson}) becomes
\begin{eqnarray}
\label{eq:Apoissonnorm}
-\Delta\hat{\vec{A}}_c&=&\frac{1}{\kappa^2}\hat{\vec{\jmath}}_s
\end{eqnarray}
Thus, it is obvious that for reasonably large values of $\kappa$, the magnetic vector potential evoked by the currents can be neglected and electromagnetic self-consistency becomes unimportant. Nevertheless, for smaller values of $\kappa$, magnetic fields become increasingly important and have to be included in the calculations.

\section{\label{sec:comp}Comparison with earlier results}

As a first test of our approach and our numerical code, we compare our results with those of Ginzburg-Landau theory presented in chapter 4.4 of Tinkham's textbook~\cite{Tin01}. There, one can find results for the critical current of a thin wire or film. In order to compare our results with these calculations, we remove the constriction ($l=0,\,w=W/2$) and consider temperatures close to $T_c$. Additionally, we have to change our boundary conditions slightly in order to allow for suppression of superconductivity in the channel at high transport current densities. This can easily be done via periodic boundary conditions at the left and right end of the channel section. After these minor changes, our program is ready to consider a long thin wire or film and our results perfectly match those from Ginzburg-Landau theory. The same results for the critical current can also be achieved by a simple Doppler shift calculation~\cite{Dah01}, i.e. solving the bulk gap and current equation with a homogeneous phase gradient
\begin{eqnarray}
i\tilde{\varepsilon}_{n}
&=&
i\varepsilon_{n}-\frac{\hbar}{2}\,\vec{v}_{F}\cdot\vec{\nabla}\phi(x)
\end{eqnarray}
This consideration provides a second verification.

As a third test, we compare our results with those for a ballistic superconducting microconstriction obtained by Zareyan et al.~\cite{Zar01}. In their work, Zareyan et al. analytically solve quasiclassical Eilenberger theory for a ballistic narrow superconducting channel connecting two bulk electrodes for temperatures close to $T_c$. If we remove the constriction from our geometry ($l=0,\,w=W/2$) and use our standard boundary conditions for the left and right section end described in section~\ref{sec:theory}, we can numerically reproduce the results given by Zareyan et al.

\section{\label{sec:jeffectinscs}The Josephson effect in ScS Junctions}

In this section, we show results without magnetic fields in order to explain in detail how the Josephson effect in a ScS junction emerges. As mentioned above, neglecting the magnetic fields caused by the supercurrents corresponds to $\kappa=\infty$.

\begin{figure}
\begin{flushleft}(a)\end{flushleft}
\vspace{-0.9cm}
\begin{center}
\includegraphics[width=0.825\columnwidth, keepaspectratio]{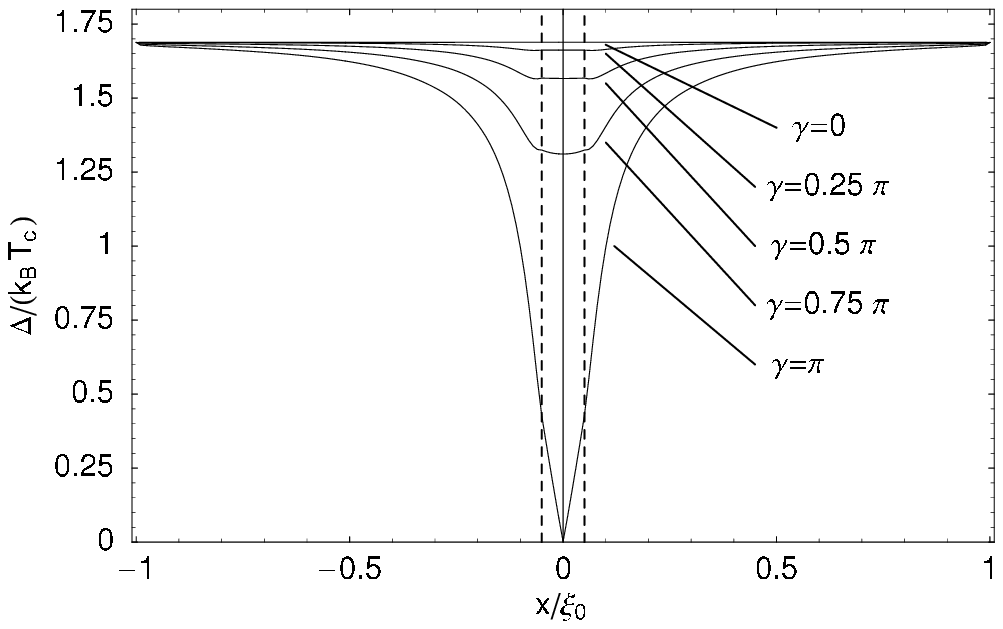}
\end{center}
\vspace{-0.5cm}
\begin{flushleft}(b)\end{flushleft}
\vspace{-0.9cm}
\begin{center}
\includegraphics[width=0.825\columnwidth, keepaspectratio]{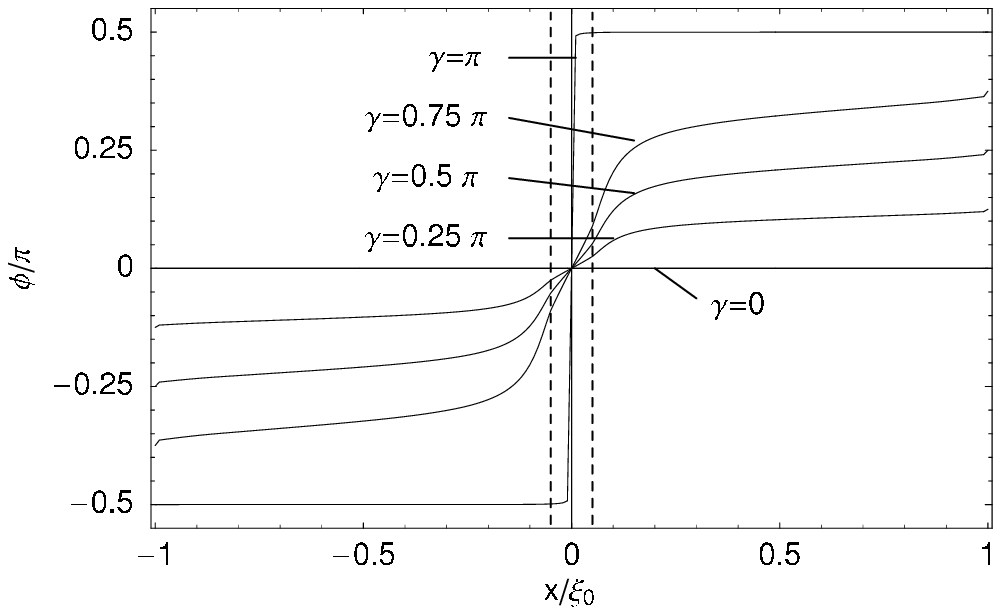}
\end{center}
\vspace{-0.5cm}
\begin{flushleft}(c)\end{flushleft}
\vspace{-0.9cm}
\begin{center}
\includegraphics[width=0.825\columnwidth, keepaspectratio]{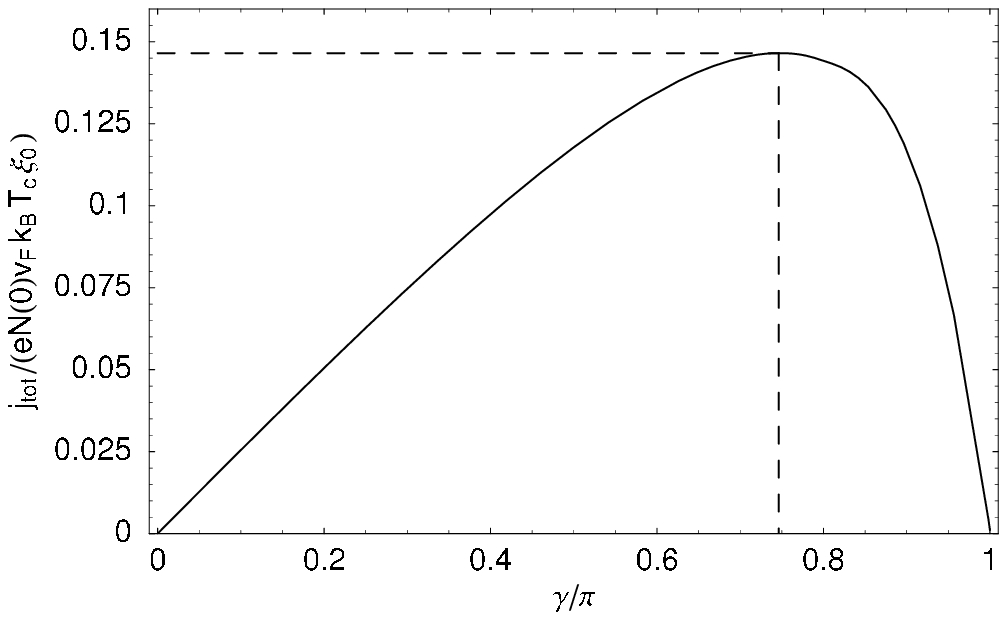}
\end{center}
\vspace{-0.5cm}
\begin{flushleft}(d)\end{flushleft}
\vspace{-0.9cm}
\begin{center}
\includegraphics[width=0.825\columnwidth, keepaspectratio]{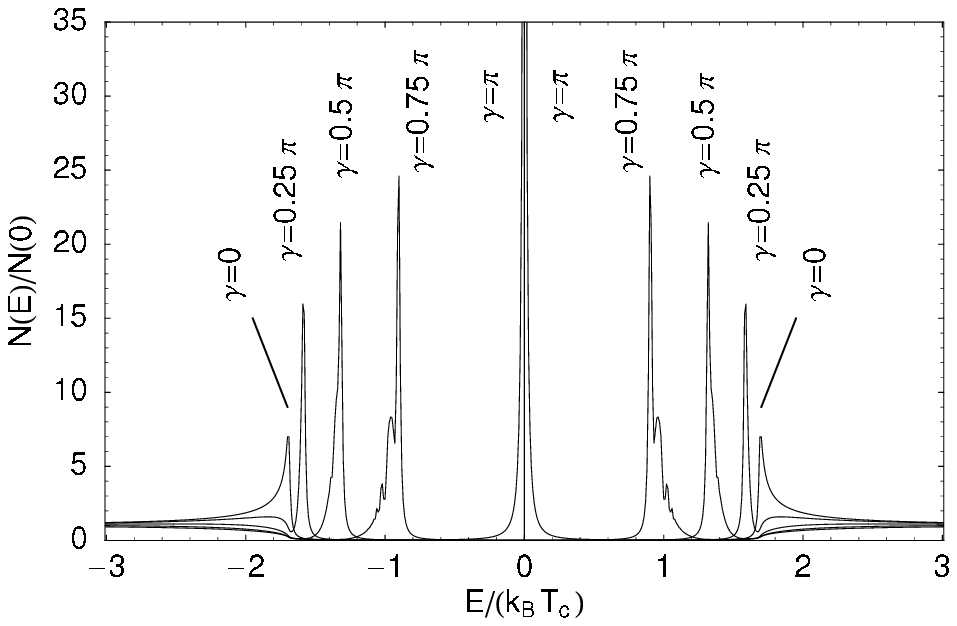}
\end{center}
\caption{\label{cap:jjindetail}Josephson effect in a ScS junction with $l=w=0.05\,\xi_0$ and $\kappa=\infty$ at a temperature of $T=0.5\,T_c$. In (a) and (b), the order parameter amplitude and phase are plotted along the $x$ axis for several values of the phase difference across the junction. The constriction length is indicated with the vertical dashed lines. In (c), we show the corresponding current-phase relation with the Josephson critical current indicated by the dashed lines. The LDOS in the center of the constriction ($x=0,y=0$) is plotted in (d).}
\end{figure}

In order to explain the Josephson effect in a ScS junction in detail, we fix the geometric parameters ($l=w=0.05\,\xi_0$) as well as the temperature ($T=0.5\,T_c$). In Fig.\,\ref{cap:jjindetail}, we show the amplitude~(a) and the phase~(b) of the order parameter for different values of the phase difference as well as the current-phase relation~(c). Additionally, we show the LDOS in the center of the constriction in~(d).

With increasing phase difference, increasing supercurrents flow across the junction. The current concentration in the constriction leads to high local current densities and causes a suppression of the order parameter amplitude. This suppression of the order parameter amplitude in turn limits the current that can flow across the junction. At the same time, the suppression of the order parameter amplitude leads to weak coupling between the left and right electrode. The suppression of the order parameter amplitude caused by the current concentration in the constriction thus evokes the Josephson effect in ScS junctions.

\begin{figure}
\begin{center}
\includegraphics[width=0.83\columnwidth, keepaspectratio]{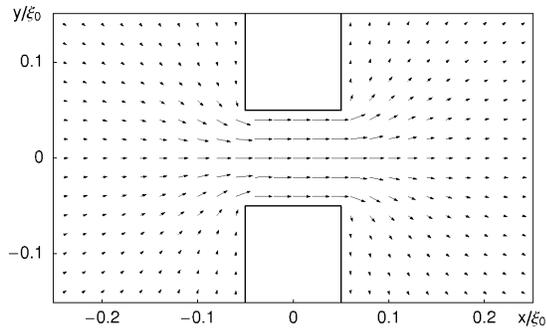}
\end{center}
\caption{\label{cap:currfieldnoemsc}Current flow pattern in the vicinity of the constriction for $l=w=0.05\,\xi_0$, $\kappa=\infty$ and $T=0.5\,T_c$ at the critical current ($\gamma=0.75\,\pi$). For better visibility, we plot only every fourth point out of the grid used in our calculations.}
\end{figure}

As long as the phase difference and the current density are small, the suppression of the order parameter amplitude is negligible and the self-consistent phase distribution resembles the solution of the Laplace equation from London theory, i.e. from local electrodynamics. The stronger the currents, the stronger is the suppression of the order parameter amplitude in the constriction and the stronger is the deviation of the phase distribution from the solution of the Laplace equation. At phase differences close to $\gamma=\pi$, the whole phase drop occurs in the constriction and only very small currents flow.
\\
For the parameters chosen here, we find that the current-phase relation strongly deviates from the standard sinusoidal form. As long as the suppression of the order parameter in the constriction is small, the current-phase relation is basically linear. However, as the suppression of the order parameter becomes stronger, the current reaches the Josephson critical current and vanishes at $\gamma=\pi$. In  Fig.\,\ref{cap:currfieldnoemsc}, we show the current flow pattern in the vicinity of the constriction at the critical current ($\gamma=0.75\,\pi$).

In Fig.\,\ref{cap:jjindetail} (d), we show the LDOS in the center of the constriction. With increasing phase difference, pronounced Andreev bound states occur at energies below the gap ($|E|<\Delta_\infty(T)$). At intermediate values of the phase difference, additional structures in the LDOS appear close to the main peak of the bound state which stem from multiple reflections in the constriction. These substructures will be discussed in detail in the next section. At the maximum phase difference of $\gamma=\pi$, the Andreev bound states occur at an energy of $E=0$ and exhibit the highest peak value.

By considering the amplitude and the phase of the order parameter, the currents and the local density of states, we gain a very complete picture of the Josephson effect in ScS junctions. We can state that the supercurrent across the constriction-type weak link is indeed mediated by the Andreev bound states. Cooper pairs incident on one side of the junction are transformed into quasiparticles that travel across the junction phase-coherently and recompose to Cooper pairs on the other side.

In order to explain the relevance of self-consistency, we compare the self-consistent results with those of a S-N-S model for the order parameter:
\begin{eqnarray}
\label{eq:SNS}
\Delta(\vec{r})
&=&
\left\{
\begin{array}{ll}
\Delta_\infty(T)\cdot e^{-i\Delta\phi/2} &, x < -l \\
0 &, |x| \leq l \\
\Delta_\infty(T)\cdot e^{+i\Delta\phi/2} &, x > +l
\end{array}
\right.
\end{eqnarray}

\begin{figure}
\begin{flushleft}(a)\end{flushleft}
\vspace{-0.9cm}
\begin{center}
\includegraphics[width=0.83\columnwidth, keepaspectratio]{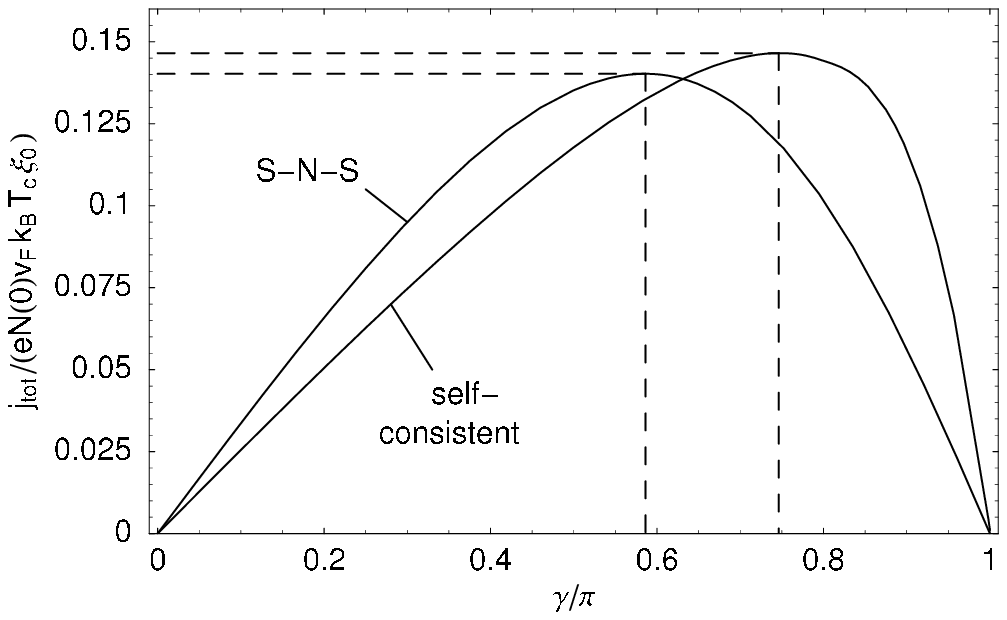}
\end{center}
\vspace{-0.5cm}
\begin{flushleft}(b)\end{flushleft}
\vspace{-0.9cm}
\begin{center}
\includegraphics[width=0.83\columnwidth, keepaspectratio]{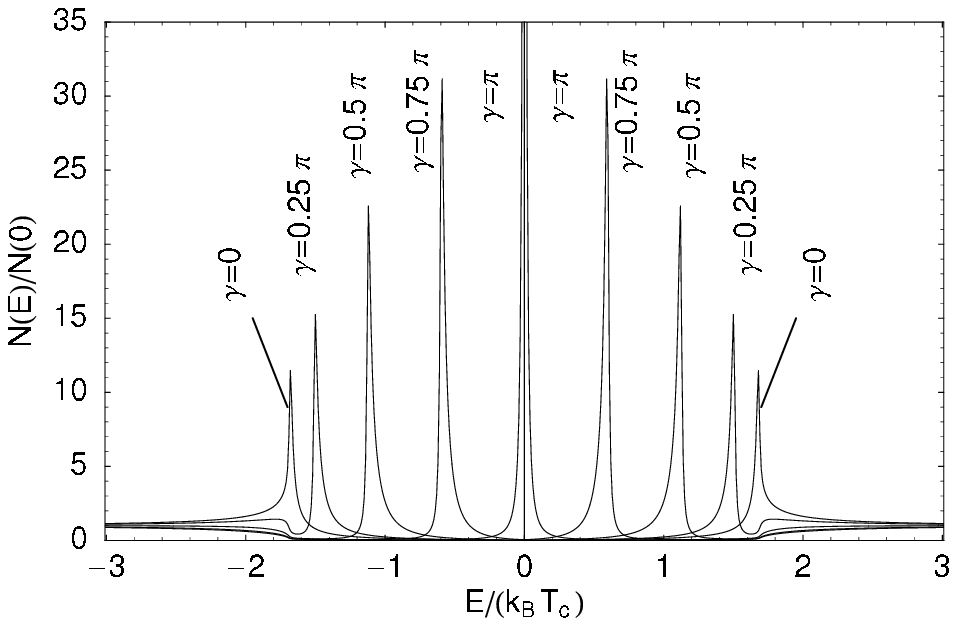}
\end{center}
\caption{\label{cap:SNS}Current-phase relation (a) and LDOS in the center of the constriction (b) for the non-self-consistent S-N-S model given in eq.\,(\ref{eq:SNS}). The self-consistent current-phase relation from Fig.\,\ref{cap:jjindetail} (c) is also plotted in (a) for comparison. For this figure, we use the same parameters as for Fig.\,\ref{cap:jjindetail}.}
\end{figure}

With this model, we calculate the supercurrents for phase differences $0\leq\gamma\leq\pi$ and extract the current-phase relation and calculate the LDOS in the center of the constriction (see Fig.\,\ref{cap:SNS}). The parameters used here are the same as for the self-consistent solutions in Fig.\,\ref{cap:jjindetail}.
\\
When using the S-N-S model, one implicitly assumes that the junction exhibits weak link behaviour. This justifies the suppression of the order parameter and allows for a rapid variation of the phase. The result is a more sinusoidal current-phase relation because the full phase drop occurs within the constriction (see Fig.\,\ref{cap:SNS} (a)). In the self-consistent solution however, part of the phase variation occurs in the channel and the effective phase difference in the direct vicinity of the constriction is smaller, at least for small total phase differences (cf. Fig.\,\ref{cap:jjindetail} (b)).
\\
The position of the bound states for the S-N-S model reflects the fact that the current-phase relation is more sinusoidal (see Fig.\,\ref{cap:SNS} (b)). Additionally, it is remarkable that the bound states in the S-N-S model do not exhibit additional substructure.

\section{\label{sec:geometry}Influence of Temperature and Geometric Parameters}

In the first part of this section, we analyze the temperature dependence of the Josephson effect in ScS junctions. In the second part, we will focus on the influence of the geometric parameters that define the constriction.

\begin{figure}
\begin{flushleft}
(a)
\end{flushleft}
\vspace{-0.9cm}
\begin{center}
\includegraphics[width=0.95\columnwidth, keepaspectratio]{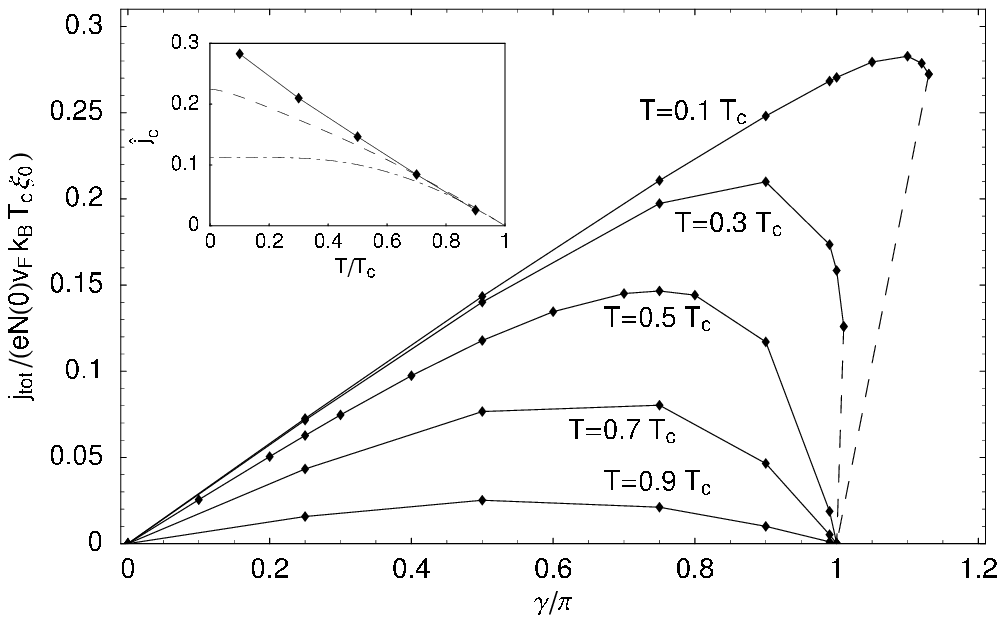}
\end{center}
\vspace{-0.5cm}
\begin{flushleft}
(b)
\end{flushleft}
\vspace{-0.9cm}
\begin{center}
\includegraphics[width=0.95\columnwidth, keepaspectratio]{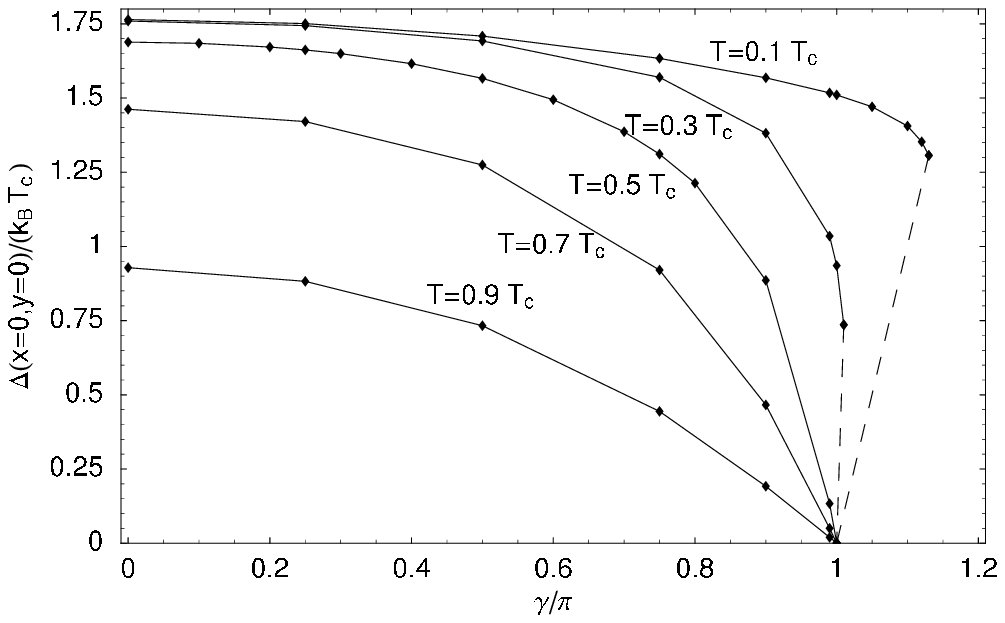}
\end{center}
\caption{\label{cap:cphirgapT}Current-phase relations for $l=w=0.05\,\xi_0$ and $\kappa=\infty$ at different temperatures are shown in (a) and the corresponding amplitude of the order parameter in the center of the constriction in (b). Points with solid guidelines are the calculated stable branches whereas the dashed lines are a rough sketch of the unstable ones. Inset in (a): Temperature-dependence of the critical current (points with solid guideline) and the Kulik-Omel'yanchuk (dashed) and the Ambegaokar-Baratoff results (dash-dotted) for comparison.}
\end{figure}

In order to investigate the temperature dependence of the Josephson effect in ScS junctions, we keep the geometric parameters fixed ($l=w=0.05\,\xi_0$) and neither consider external magnetic fields nor those caused by the supercurrents ($\hat{B}_0=0$, $\kappa=\infty$).
\\
To study the influence of the temperature, we calculate the order parameter $\Delta(\vec{r})$, the currents $\vec{\jmath}(\vec{r})$ and the local density of states $N(E)$ for different values of the phase difference $\gamma$ at temperatures from $T=0.1\,T_c$ to $0.9\,T_c$. From the resulting current configurations, we extract the current-phase relations which we show in Fig.\,\ref{cap:cphirgapT} (a). In Fig.\,\ref{cap:cphirgapT} (b), we plot the amplitude of the order parameter in the center of the constriction for the same values of $T$ and $\gamma$ as for the current-phase relations. For the two highest temperatures, we increase the cutoff $\omega_c$ used in the gap equation and the current equation in order to include a sufficient number of Matsubara frequencies ($\omega_c=100\,k_B T_c$ for $T=0.7\,T_c$ and $\omega_c=200\,k_B T_c$ for $T=0.9\,T_c$ respectively).

At temperatures close to $T_c$, the current-phase relation is highly sinusoidal. The lower the temperature, the stronger are the deviations of the current-phase relation from $\sin(\gamma/2)$. For the geometry used here, the current-phase relation becomes multivalued at temperatures $T\leq 0.3\,T_c$. As stated above, our iterative approach yields only the stable branch of the solution. The unstable branch of the current-phase relation is roughly sketched in the diagram (dashed lines, see e.g.~\cite{Mar01,Lev01}). The multivalued character of the current-phase relation implies that the current increases even for phase differences $\gamma>\pi$ and  marks the transition from Josephson junction behaviour to bulk current flow.
\\
In the inset of Fig.\,\ref{cap:cphirgapT} (a), we plot the dependence of the critical current on temperature (points with solid guideline). For comparison, we plot the Kulik-Omel'yanchuk result from~\cite{Kul02} for a point contact in the clean limit (dashed line) as well as the Ambegaokar-Baratoff result for a tunnel junction (dash-dotted) from~\cite{Amb01}. Close to $T_c$, the critical current exhibits a linear decrease with increasing temperature and the self-consistent results of our calculations coincide with those for the point contact. At lower temperatures however, the self-consistent treatment leads to a stronger increase of the critical current than predicted for the point contact. In the Kulik-Omel'yanchuk results for the point contact, the current vanishes at $\gamma=\pi$ for all temperatures whereas, in the self-consistent treatment, the currents increase for $\gamma>\pi$~\cite{Mar01,Lev01}. At low temperatures, this leads to higher values of the critical current in the self-consistent calculation. In calculations for a more point contact-like constriction ($l=w=0.01\xi_0$), we find that this correction due to self-consistency remains.

\begin{figure}
\begin{center}
\includegraphics[width=1\columnwidth, keepaspectratio]{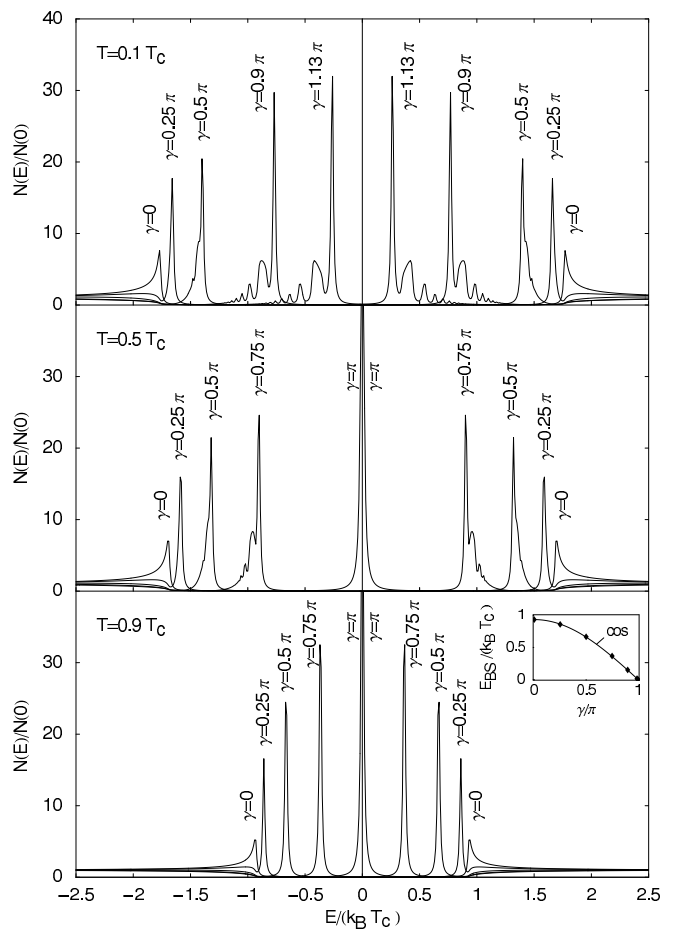}
\end{center}
\caption{\label{cap:LDOST}Local density of states in the center of the constriction at three different temperatures ($T=0.1\,T_c$, $0.5\,T_c$ and $0.9\,T_c$ from top to bottom) for different values of the phase difference $\gamma$ as indicated in the diagrams. Inset: Position of the Andreev bound states at $T=0.9\,T_c$ as a function of the phase difference $\gamma$ (dots) and $E=\Delta_\infty(T)\cos(\gamma / 2)$ for comparison. For this figure, we use the same parameters as for Fig.\,\ref{cap:cphirgapT}.}
\end{figure}

\begin{figure}
\begin{flushleft}
(a)
\end{flushleft}
\vspace{-0.9cm}
\begin{center}
\includegraphics[width=0.95\columnwidth, keepaspectratio]{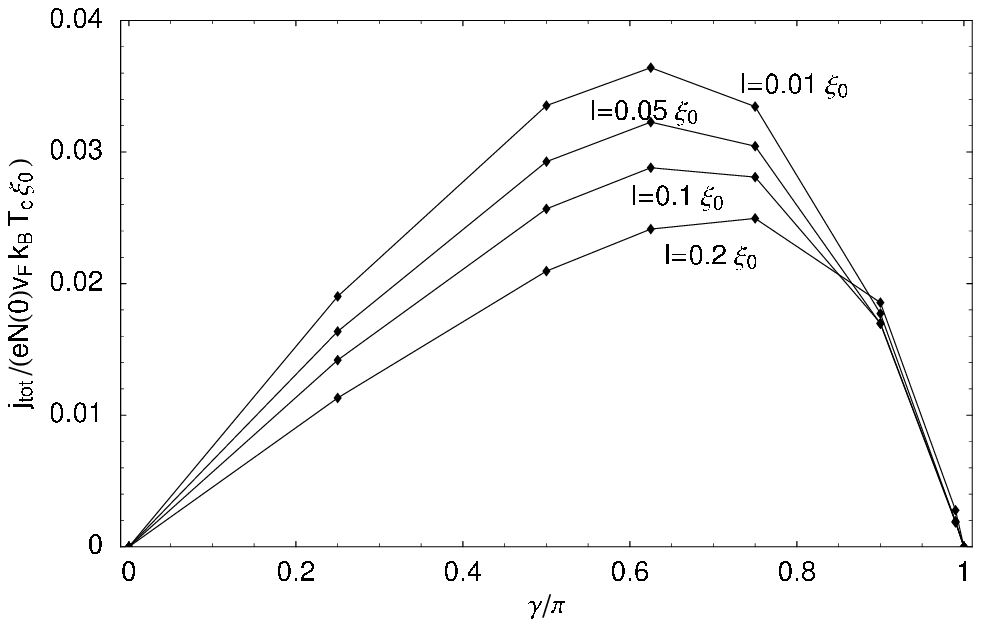}
\end{center}
\vspace{-0.5cm}
\begin{flushleft}
(b)
\end{flushleft}
\vspace{-0.9cm}
\begin{center}
\includegraphics[width=0.95\columnwidth, keepaspectratio]{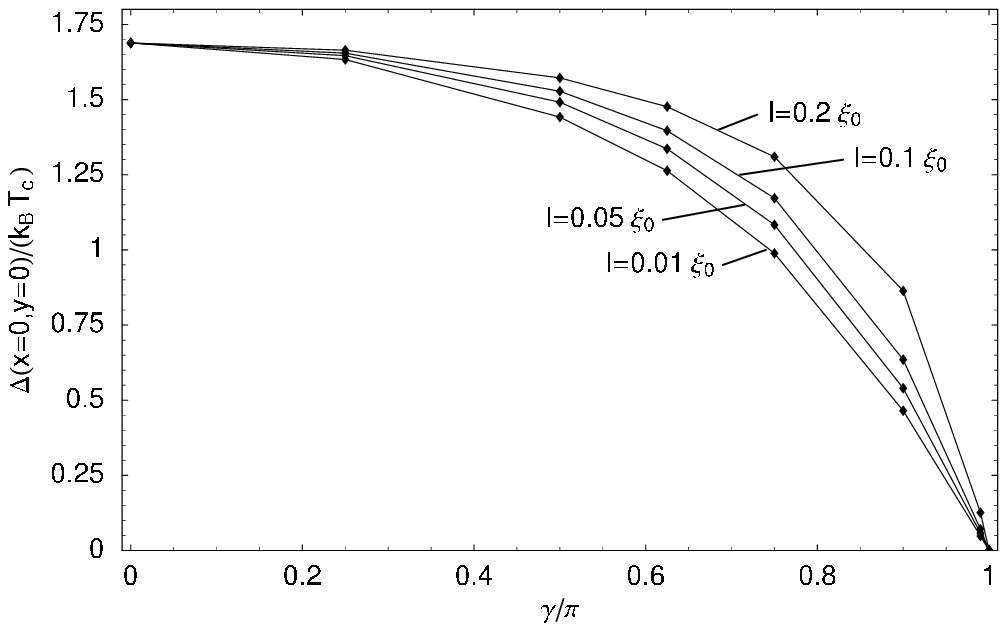}
\end{center}
\caption{\label{cap:cphirgapll}Current-phase relation (a) and order parameter amplitude in the center of the constriction (b) for different values of the constriction length $l$ as indicated in the diagrams. For this figure, we use $T=0.5\,T_c$, $\kappa=\infty$ and $w=0.01\,\xi_0$ and do not consider magnetic fields. Lines are guides for the eye.}
\end{figure}

In Fig.\,\ref{cap:LDOST}, we show the local density of states in the center of the constriction for three different temperatures. 
With increasing phase difference and thus increasing transport current, Andreev bound states occur in the constriction. The larger the phase difference, the lower is the energy of the bound states. At $T=0.1\,T_c$, the Andreev bound state for $\gamma=\pi$ still has finite energy. At higher temperatures, when the current-phase relation is not multivalued, the bound states have zero energy for the maximum phase difference $\gamma=\pi$.
\\
For $T=0.9\,T_c$, we additionally plot the position of the Andreev bound states as a function of the phase difference $\gamma$ (inset in lowermost panel of Fig.\,\ref{cap:LDOST}). At temperatures close to the critical temperature $T_c$, the position of the bound states is given by $E=\pm\Delta_\infty(T)\cos(\gamma / 2)$ which is plotted in the inset for reference. At lower temperatures, deviations occur as can be seen in the upper two panels of Fig.\,\ref{cap:LDOST}.
\\
Primarily at low temperatures, the Andreev bound states do not only have one peak but exhibit additional local maxima at energies above the main peak. This substructure is due to multiple reflections of quasiparticles within the constriction.
If we consider the center of the constriction and measure the angle of the trajectories $\alpha$ with respect to the $x$ axis, an additional reflection within the constriction sets in at angles $\alpha_n$ with
\begin{eqnarray*}
\tan\left( \alpha_n \right)
&=&
(2n+1)\cdot\frac{w}{l},
\hspace{0.2cm}
n=0,1,2,\dots
\end{eqnarray*}
The intervals between the angles $\alpha_n$ each correspond to a local maximum of the bound state. The effective distance over which the phase drop occurs along the trajectory increases with every additional reflection and the effective phase gradient thus decreases.
\\
In the case of the S-N-S model, the substructure of the bound states does not occur because the phase drop occurs in the direct vicinity of the constriction. Reflections of the trajectories do not change the effective distance over which the phase drop occurs along the trajectories.

\begin{figure}
\begin{flushleft}
(a)
\end{flushleft}
\vspace{-0.9cm}
\begin{center}
\includegraphics[width=0.95\columnwidth, keepaspectratio]{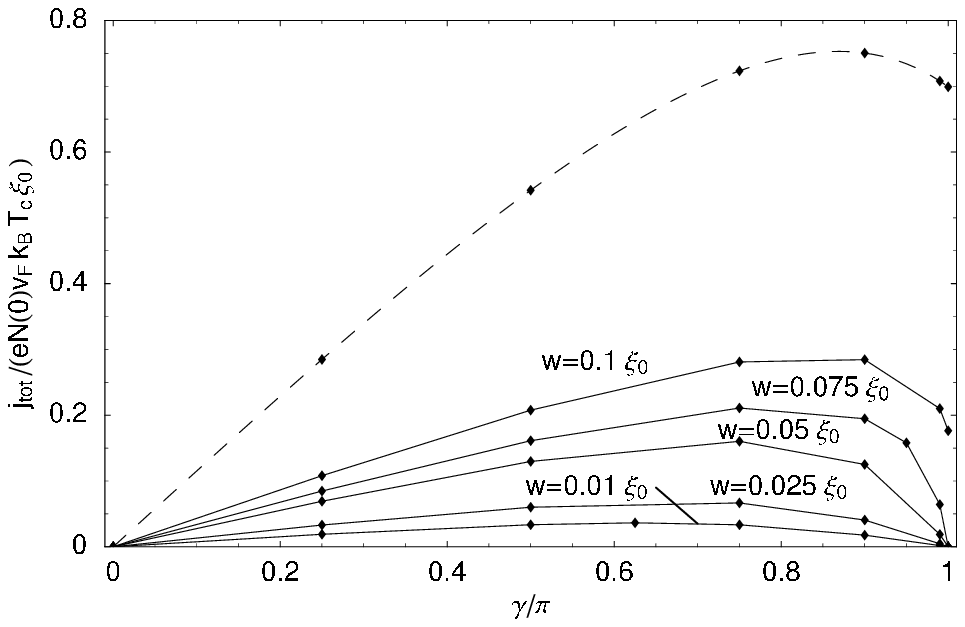}
\end{center}
\vspace{-0.5cm}
\begin{flushleft}
(b)
\end{flushleft}
\vspace{-0.9cm}
\begin{center}
\includegraphics[width=0.95\columnwidth, keepaspectratio]{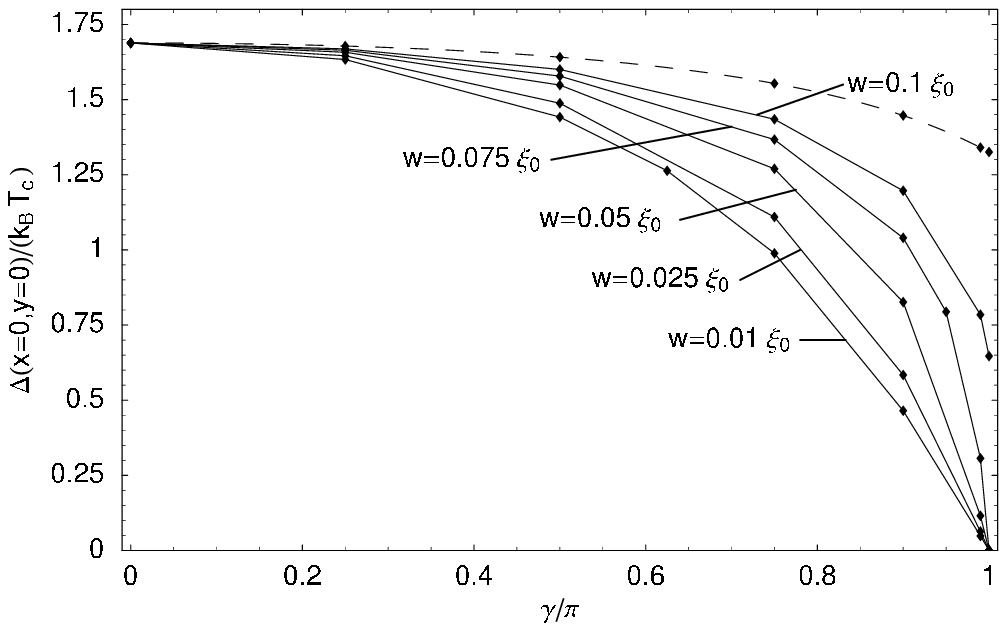}
\end{center}
\caption{\label{cap:cphirgapbb}Current-phase relation (a) and order parameter amplitude in the center of the constriction (b) for different values of the constriction width $w$ as indicated. The dashed lines correspond to a channel without constriction. For this figure, we use $T=0.5\,T_c$, $\kappa=\infty$ and $l=0.01\,\xi_0$ and do not consider magnetic fields. Lines are guides for the eye.}
\end{figure}

Now let us address the influence of the geometric parameters on the Josephson effect.
\\
In Fig.\,\ref{cap:cphirgapll}, we plot the current-phase relation and the order parameter amplitude in the center of the constriction for different values of the constriction length $l$. The temperature is set to $T=0.5\,T_c$, the width of the constriction to $w=0.01\,\xi_0$ and we do not consider magnetic fields. If the constriction is shorter, more current flows and the suppression of the order parameter amplitude is stronger. Thus, for shorter constrictions, the current-phase relation is more sinusoidal and the Josephson critical current is higher than for longer ones. The decrease of the critical current with increasing length of the constriction is consistent with the results found by Zareyan et al.~\cite{Zar01} for temperatures close to $T_c$.

In Fig.\,\ref{cap:cphirgapbb}, we show the current-phase relation and the order parameter amplitude in the center of the constriction for different values of the constriction width $w$. As before, the temperature is set to $T=0.5\,T_c$, the length of the constriction to $l=0.01\,\xi_0$ and we do not consider magnetic fields. For reference, we show the current-phase relation without a constriction but with periodic boundary conditions as described in section \ref{sec:comp} (dashed lines in both Fig.\,\ref{cap:cphirgapbb} (a) and (b)). The dashed lines thus correspond to a long superconducting lead with homogeneous current flow or equally a homogenous bulk current.

As expected, the width of the constriction strongly influences the Josephson critical current. The narrower the constriction, the stronger is the current concentration in the constriction and the stronger is the suppression of the order parameter amplitude. Hence, less current flows at the same phase difference for a narrower constriction.

\section{\label{sec:kappa}Effects of Screening}

From this section on, we will include the effects of magnetic fields in our considerations. In this section, we will focus on the influence of the parameter $\kappa=\lambda_L/\xi_0$ whereas in the next section, we will examine the effect of an external magnetic field.

So far, we neither considered external magnetic fields nor those caused by the supercurrents. Neglecting the latter corresponds to large values of $\kappa$ which can easily be seen from equation~(\ref{eq:Apoissonnorm}). In our case, the lateral extension of the geometry is $W=1\,\xi_0$ and thus not much happens for $\kappa\gtrsim 1$. Nevertheless, for smaller values of $\kappa$, the distribution of the currents and the magnetic fields changes drastically.

\begin{figure}
\begin{flushleft}
(a)
\end{flushleft}
\vspace{-0.9cm}
\begin{center}
\includegraphics[width=0.93\columnwidth, keepaspectratio]{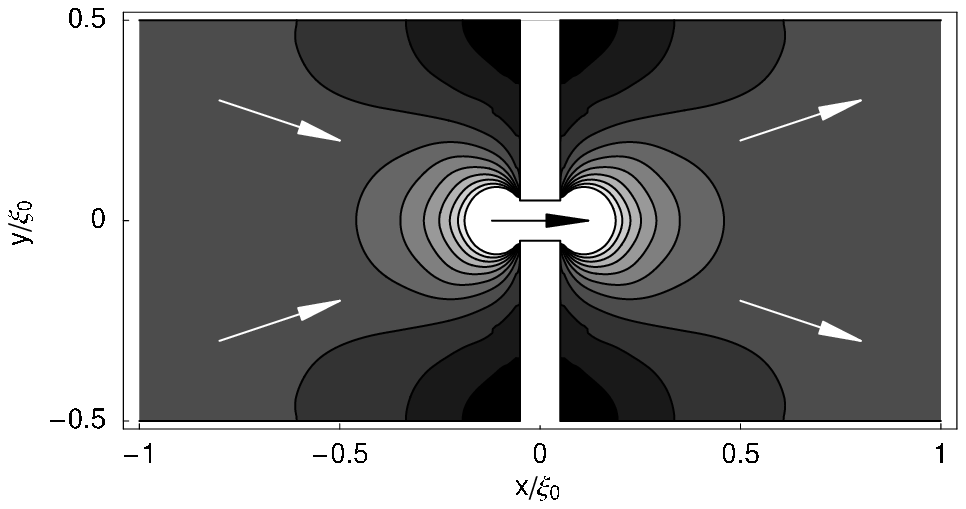}
\end{center}
\vspace{-0.5cm}
\begin{flushleft}
(b)
\end{flushleft}
\vspace{-0.9cm}
\begin{center}
\includegraphics[width=0.93\columnwidth, keepaspectratio]{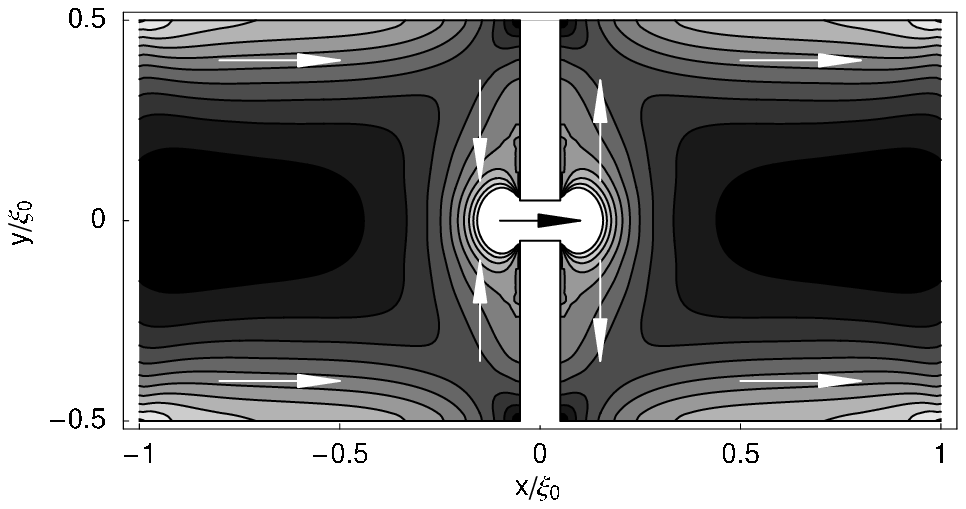}
\end{center}
\caption{\label{cap:currdensityGLk}Current density at the critical current for $\kappa=1$ (a) and $\kappa=0.1$ (b). Dark shading represents low current density wheres bright areas exhibit high current density. Arrows indicate the direction of the current flow. For small values of $\kappa$, the current is strongly concentrated on the surface of the sample.}
\end{figure}

In Fig.\,\ref{cap:currdensityGLk}, we show the current density at the Josephson critical current for $\kappa=1$ (a) and $\kappa=0.1$ (b). To calculate these configurations, we use $T=0.5\,T_c$ and $l=w=0.05\,\xi_0$. For $\kappa=1$, the current density is homogeneous over the cross section of the channel. Nevertheless, in the vicinity of the constriction, the current is strongly concentrated. For $\kappa=0.1$, the current is concentrated on the surface of the sample. In the channel, the current distribution thus changes drastically. However, since the constriction size is equal to the London penetration depth $\lambda_L=\kappa\,\xi_0$ with $\kappa=0.1$ used here, the current density is homogeneous in the constriction.

\begin{figure}
\begin{flushleft}
(a)
\end{flushleft}
\vspace{-0.9cm}
\begin{flushright}
\includegraphics[width=0.93\columnwidth, keepaspectratio]{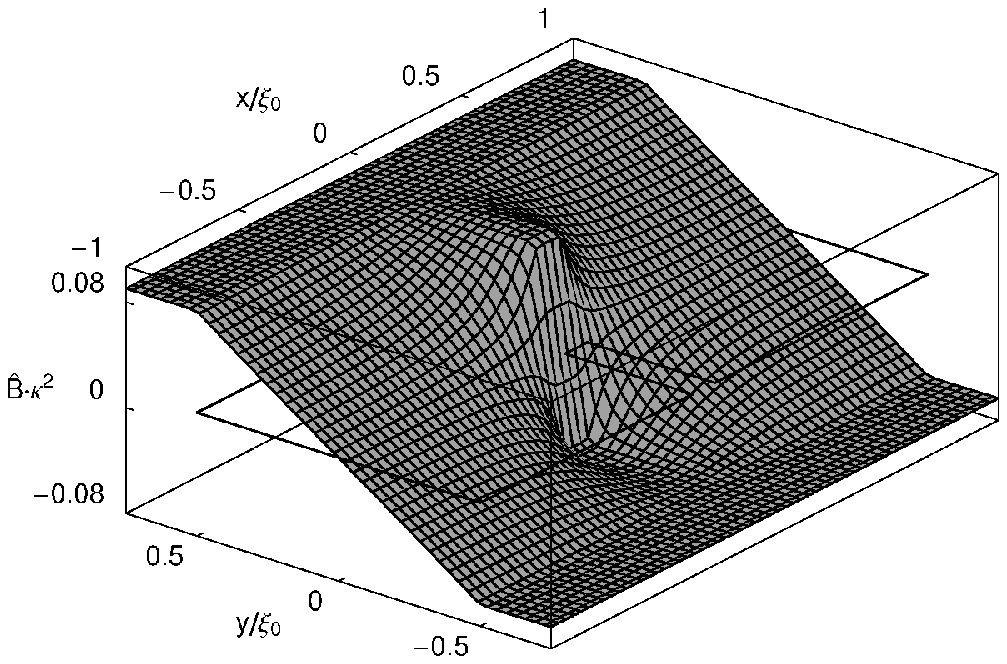}
\end{flushright}
\vspace{-0.5cm}
\begin{flushleft}
(b)
\end{flushleft}
\vspace{-0.9cm}
\begin{flushright}
\includegraphics[width=0.93\columnwidth, keepaspectratio]{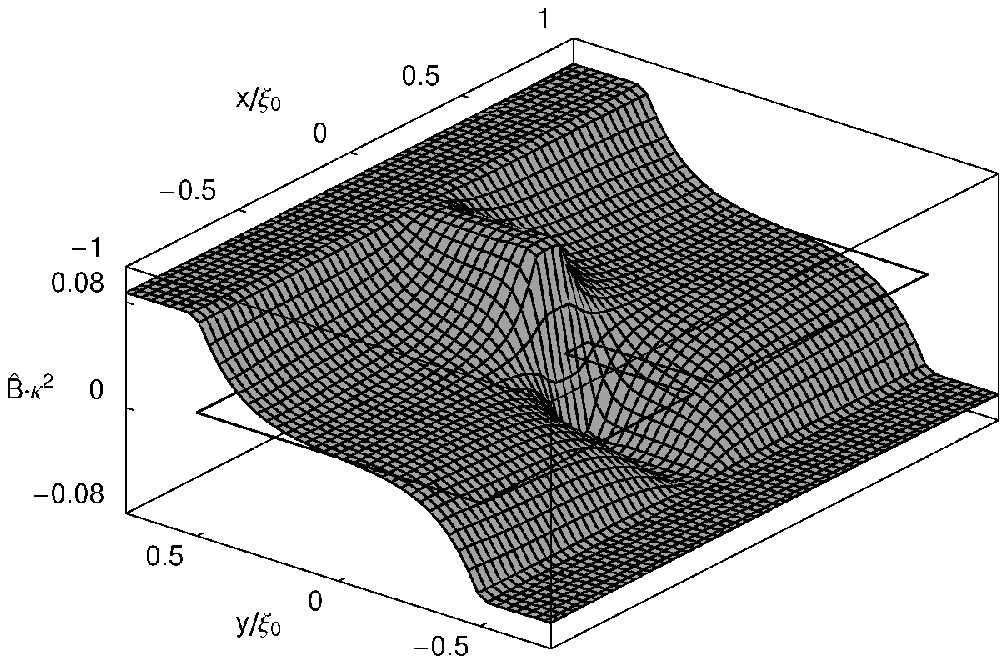}
\end{flushright}
\caption{\label{cap:magfieldGLk}Magnetic field distribution $\hat{\vec{B}}=\hat{B}\,\hat{e}_z$ for $\kappa=1$ (a) and $\kappa=0.1$ (b). We multiply the magnetic field $\hat{B}$ with $\kappa^2$ in order to have equal scaling of the vertical axis. The geometry of the sample is indicated with the black lines. For smaller values of $\kappa$, the current is concentrated on the surface of the sample and the magnetic field is suppressed in the interior.}
\end{figure}

In Fig.\,\ref{cap:magfieldGLk}, we plot the magnetic fields corresponding to the current configurations in Fig.\,\ref{cap:currdensityGLk}. The strong concentration of the current on the surface of the sample leads to the disappearance of the magnetic field in the interior of the channel. Nevertheless, the magnetic field in the vicinity of the constriction does not depend much on the value of $\kappa$ as long as the dimensions of the constriction are smaller than the London penetration depth ($\lambda_L\gtrsim l,w$). Accordingly, the current-phase relation and the dependence of the order parameter amplitude in the center of the constriction on the phase difference remain unchanged for $\kappa\gtrsim 0.1$. Thus, we can conclude that the Josephson effect is independent of $\kappa$ as long as $\lambda_L\gtrsim l,w$.

\begin{figure}
\begin{center}
\includegraphics[width=1\columnwidth, keepaspectratio]{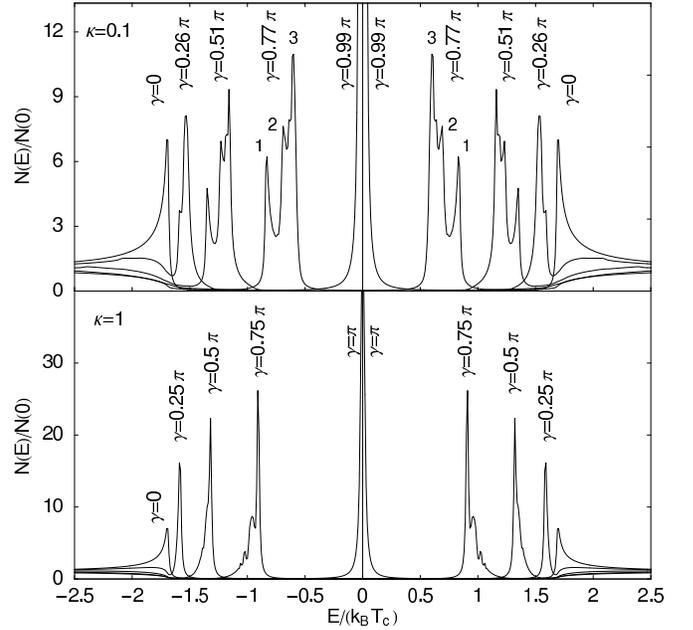}
\end{center}
\caption{\label{cap:ldosGLk}Local density of states in the center of the constriction for two different values of $\kappa$ ($\kappa=0.1$ in the upper panel, $\kappa=1$ in the lower panel). For this figure, we use $T=0.5\,T_c$ and $l=w=0.05\,\xi_0$. Please note the different scaling of the vertical axis for the two panels.}
\end{figure}

In Fig.\,\ref{cap:ldosGLk}, we show the local density of states in the center of the constriction for $T=0.5\,T_c$ and $l=w=0.05\,\xi_0$. The inhomogeneous configuration for small values of $\kappa$ leads to a splitting of the bound states. For the case of $\kappa=0.1$ and $\gamma=0.77\,\pi$, we discuss the origin of the peaks labeled 1,2,3 in the upper panel of Fig.\,\ref{cap:ldosGLk} in detail.
\\
Peak 1 originates from trajectories which are oriented parallel to the $x$ axis and thus pass through regions with small current density. Peaks 2 and 3 stem from trajectories which are reflected into the lower ($y<0$) and upper ($y>0$) part of the channel, i.e. in regions with high current density. Thus, the concentration of the supercurrents on the surface of the sample leads to differing contributions of the trajectories for different angles and yields a splitting of the bound state.

\section{\label{sec:bext}Influence of the External Magnetic Field}

In this section, we will finally examine the influence of an external magnetic field on the Josephson effect in a ScS junction. Therefore, we set the temperature to $T=0.5\,T_c$, the geometric parameters to $l=w=0.05\,\xi_0$, the parameter $\kappa=1$ and apply an external magnetic field $\hat{\vec{B}}_0=\hat{B}_0\,\hat{e}_z$. At the end of this section, we will discuss the temperature dependence of the influence of the external magnetic field.

Before turning to the results of our calculations, let us briefly discuss the normalization we use for the magnetic field in terms of more familiar quantities. Following the procedure presented in~\cite{Dah03}, we calculate the bulk upper critical field $B_{c2}$ using Eilenberger theory~\cite{Eil02}. For s-wave superconductors with cylindrical Fermi surface and with the magnetic field oriented parallel to the cylinder axis, we find
\begin{eqnarray*}
\ln\left( \frac{T}{T_c}\right)
&=&
\int_0^\infty \frac{du}{\sinh u}\cdot
\left(
e^{-u^2 \cdot \frac{1.764}{8\pi^2} \cdot \hat{B}_{c2} \cdot \left(\frac{T_c}{T}\right)^2} - 1
\right)
\end{eqnarray*}
At $T=0.5\,T_c$, the normalized upper critical field thus is $\hat{B}_{c2}=4.51$, or $B_{c2}=0.424\cdot\big( -\frac{dB_{c2}}{dT}\big|_{T_c}\cdot T_c \big)$ respectively.

\begin{figure}
\begin{flushleft}
(a)
\end{flushleft}
\vspace{-0.9cm}
\begin{center}
\includegraphics[width=0.93\columnwidth, keepaspectratio]{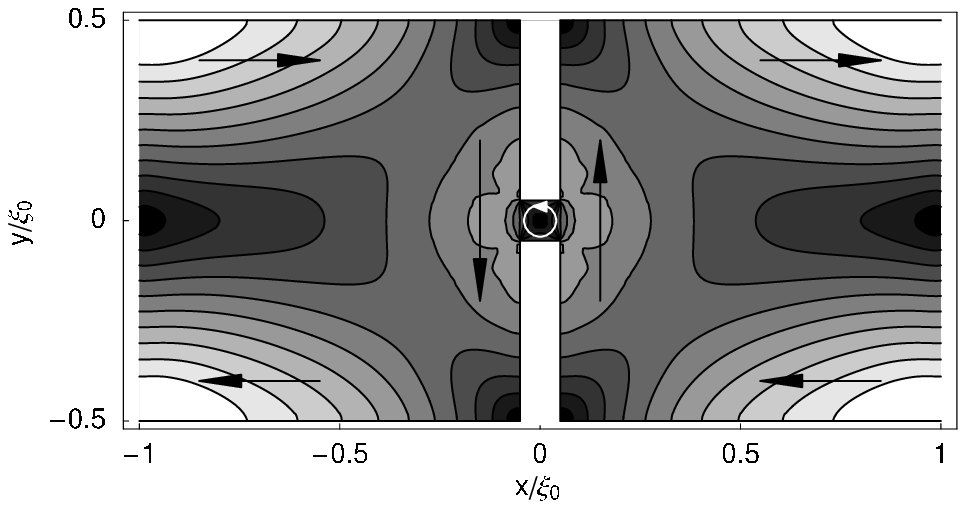}
\end{center}
\vspace{-0.5cm}
\begin{flushleft}
(b)
\end{flushleft}
\vspace{-0.9cm}
\begin{center}
\includegraphics[width=0.93\columnwidth, keepaspectratio]{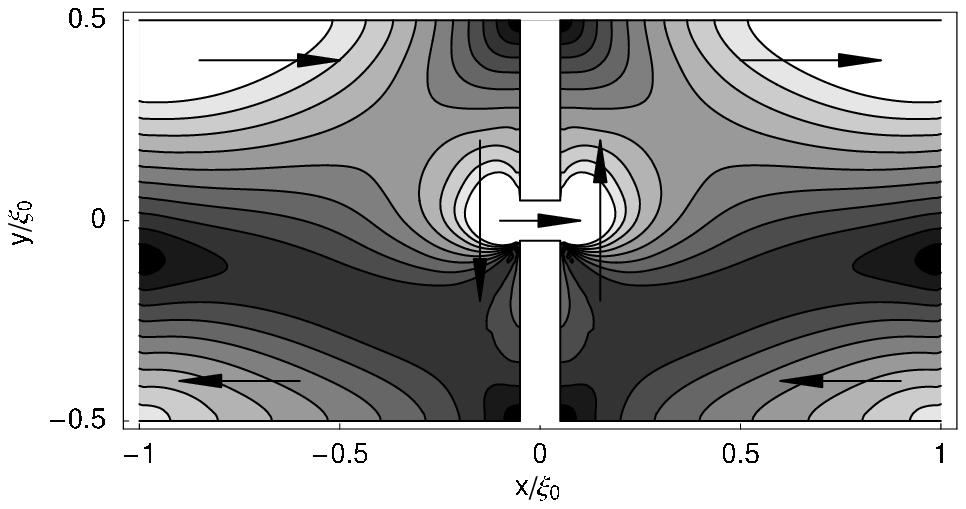}
\end{center}
\caption{\label{cap:currdensityB0}Current density with applied external magnetic field ($\kappa=1$, $\hat{B}_0=2$) without transport current ($\gamma=0$, (a)) and with critical transport current ($\gamma=0.74\,\pi$, (b)). Again, dark shading signifies low current density whereas bright shading signifies high current density. Arrows indicate the direction of the current flow. Without transport currents, a small circulating current appears in the constriction which leads to local antiscreening (white circular arrow in the constriction in (a)). With an applied phase difference, the total current distribution becomes asymmetric and consists of both screening and transport currents.}
\end{figure}

\begin{figure}
\begin{flushleft}
(a)
\end{flushleft}
\vspace{-0.9cm}
\begin{flushright}
\includegraphics[width=0.93\columnwidth, keepaspectratio]{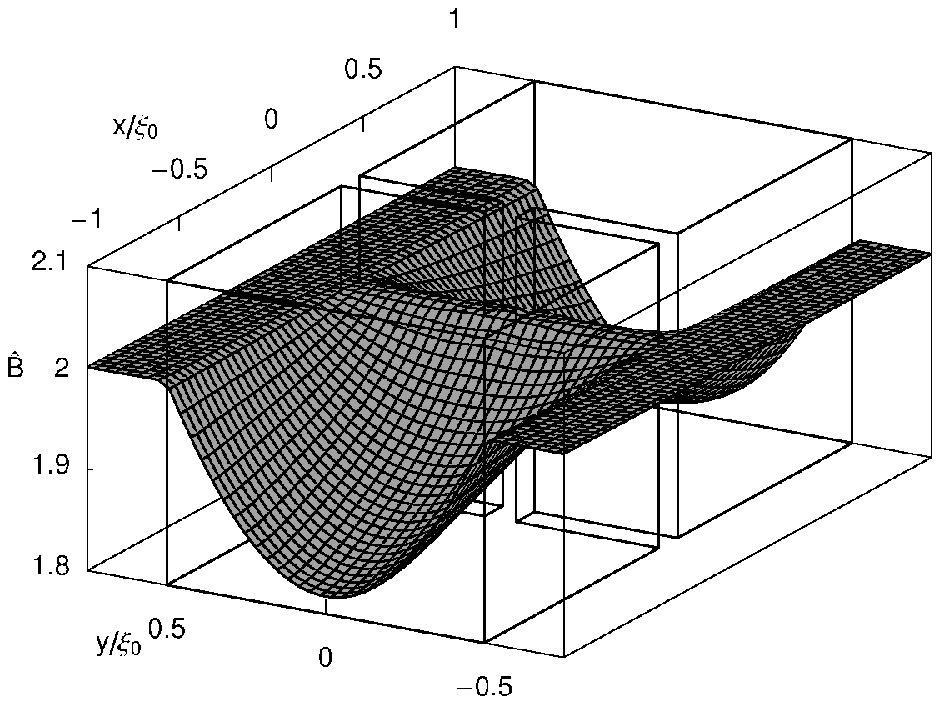}
\end{flushright}
\vspace{-0.5cm}
\begin{flushleft}
(b)
\end{flushleft}
\vspace{-0.9cm}
\begin{flushright}
\includegraphics[width=0.93\columnwidth, keepaspectratio]{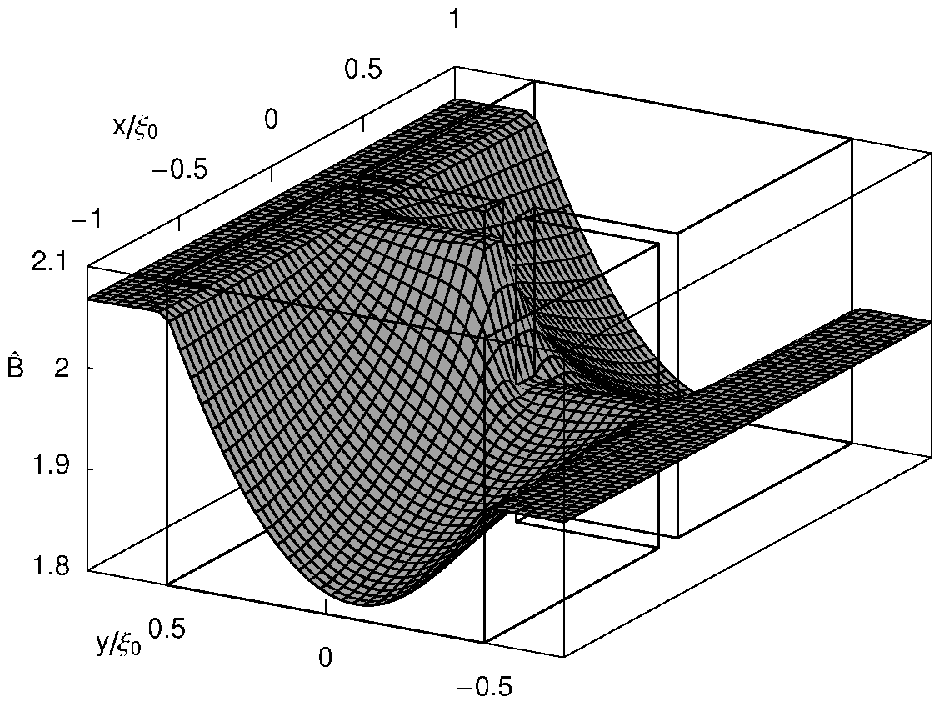}
\end{flushright}
\caption{\label{cap:magfieldB0}Magnetic field distribution $\hat{\vec{B}}=\hat{B}\,\hat{e}_z$ for applied external magnetic field ($\kappa=1$, $\hat{B}_0=2$) without transport current ($\gamma=0$, (a)) and with critical transport current ($\gamma=0.74\,\pi$, (b)). The geometry of the sample is indicated with the black lines. Screening currents lead to a reduction of the magnetic field in the interior of the sample. Additional transport currents increase the magnetic field on one side of the sample and decrease it on the other side.}
\end{figure}

An external magnetic field leads to screening currents which flow on the surface of the superconductor. Thus, in our case, the currents are a mixture of transport currents across the Josephson junction and screening currents. In Fig.\,\ref{cap:currdensityB0}, we show the current density in the section enclosing the constriction for an external magnetic field of $\hat{B}_0=2$ without transport currents ($\gamma=0$, (a)) and with critical transport currents ($\gamma=0.74\,\pi$, (b)). In Fig.\,\ref{cap:magfieldB0}, we plot the corresponding magnetic field distributions $\hat{B}\,\hat{e}_z$.

Without transport currents, we find screening currents which flow on the surface of the channel and which lower the magnetic field inside the sample. Since the width of the channel is equal to the London penetration depth ($\lambda_L=\kappa\,\xi_0$ with $\kappa=1$ used here), the magnetic field is not completely suppressed. In the constriction, a small circulating current flows which is directed opposite to the screening currents and which leads to a local enhancement or antiscreening of the magnetic field.
\\
With a finite phase difference applied between the ends of the section, a transport current flows. This leads to a breaking of the fourfold symmetry visible in Fig.\,\ref{cap:currdensityB0}~(a) and to a combined current configuration consisting of screening as well as transport currents. Consequently, the magnetic field distribution shows an asymmetry. The total magnetic field is decreased for $y<0$ and increased for $y>0$. Far away from the sample, the external magnetic field is recovered, but this happens on a much larger scale and can not be seen in Fig.\,\ref{cap:magfieldB0}.

\begin{figure}
\begin{flushleft}
(a)
\end{flushleft}
\vspace{-0.9cm}
\begin{center}
\includegraphics[width=0.95\columnwidth, keepaspectratio]{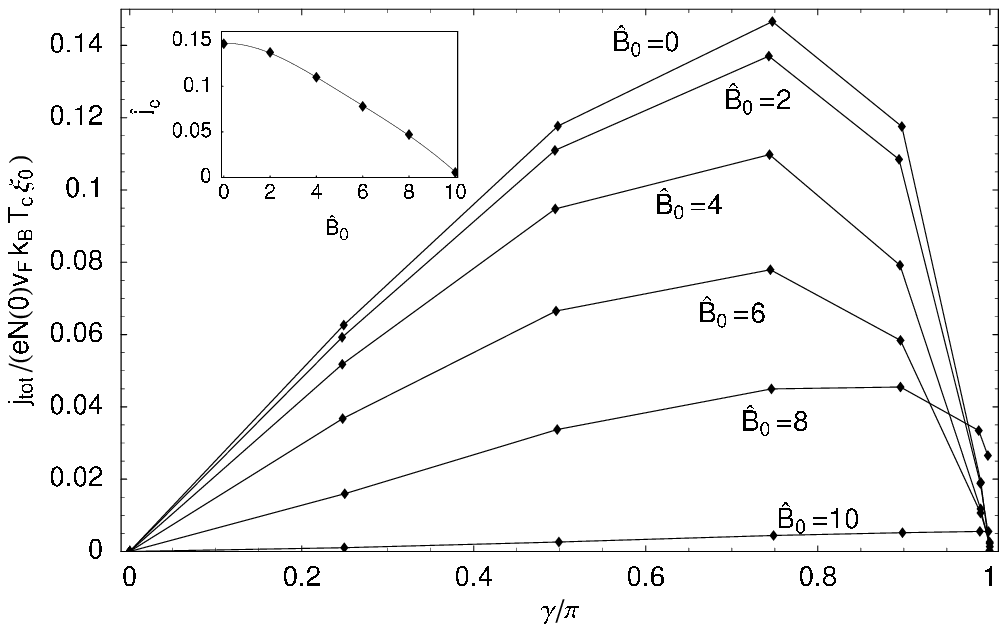}
\end{center}
\vspace{-0.5cm}
\begin{flushleft}
(b)
\end{flushleft}
\vspace{-0.9cm}
\begin{center}
\includegraphics[width=0.95\columnwidth, keepaspectratio]{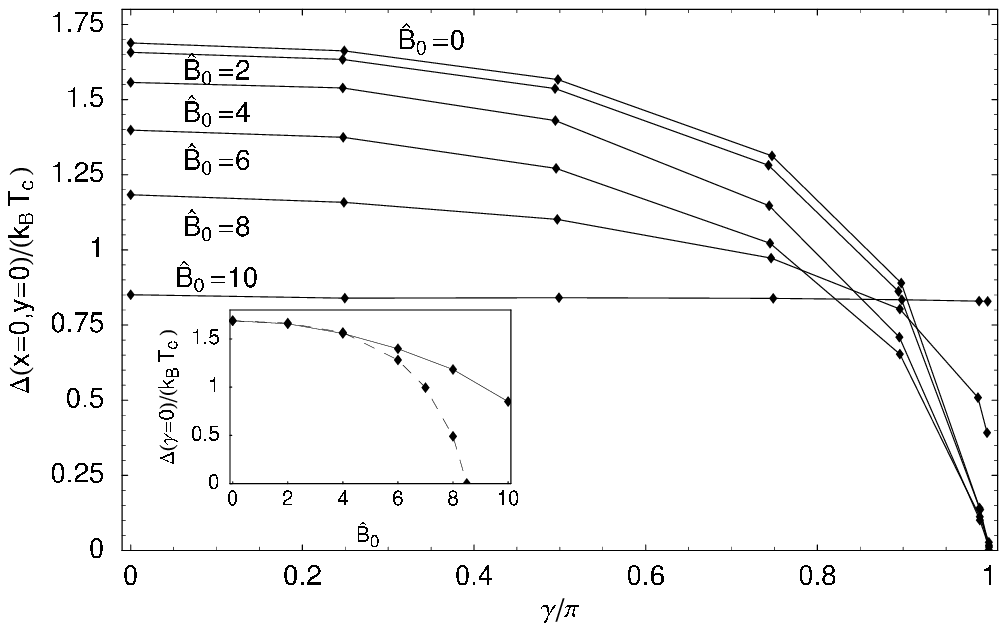}
\end{center}
\caption{\label{cap:cphirgapB0}Current-phase relation (a) and order parameter amplitude in the center of the constriction (b) for different values of the external magnetic field $\hat{B}_0$. Inset in (a): Dependence of the Josephson critical current on $\hat{B}_0$. Inset in (b): Amplitude of the order parameter in the center of the constriction (solid line) and in the center of a long channel without constriction (dashed line) for $\gamma=0$ as a function of the external magnetic field. For the whole figure, we use $T=0.5\,T_c$, $l=w=0.05\,\xi_0$ and $\kappa=1$. Lines are guides for the eye.}
\end{figure}

In Fig.\,\ref{cap:cphirgapB0}, we show the current-phase relation and the order parameter amplitude in the center of the constriction for different values of the external magnetic field. For small values of the external magnetic field, the strength of the currents is gradually reduced as is the order parameter amplitude in the constriction, but the functional form of the current-phase relation and the order parameter amplitude is basically unchanged. At larger values of the external magnetic field however, the transport currents become very small and the order parameter amplitude in the constriction remains unaffected. At the same time, the Josephson effect disappears.

The reason for the disappearance of the Josephson effect is that with increasing external magnetic field, superconductivity is suppressed in the channel. Since the constriction is much smaller than the London penetration depth, the screening currents do not flow across the constriction (see Fig.\,\ref{cap:currdensityB0} (a)). This leads to pair breaking in the channel and lowers the amplitude of the order parameter. The screening currents mainly flow in the channel and superconductivity first breaks down here. However, if superconductivity is already suppressed in the channel, no current can flow across the constriction and the Josephson effect does not occur.

The external magnetic field which leads to suppression of the Josephson effect is about twice the approximate upper critical field $\hat{B}_{c2}$ derived above. To verify the validity of this unusually large value of the magnetic field leading to the suppression of superconductivity in the channel, we carried out calculations for a longer channel with the same width ($L=10\,\xi_0$, $W=1\,\xi_0$), but without a constriction. The amplitude of the order parameter in the center of the constriction as well as in the center of the long channel for $\gamma=0$ is plotted in the inset in Fig.\,\ref{cap:cphirgapB0} (b). In the channel, superconductivity is completely suppressed at somewhat lower values of the external magnetic field, namely at about $\hat{B}_0\approx 8.5$.

\begin{figure}
\begin{center}
\includegraphics[width=1\columnwidth, keepaspectratio]{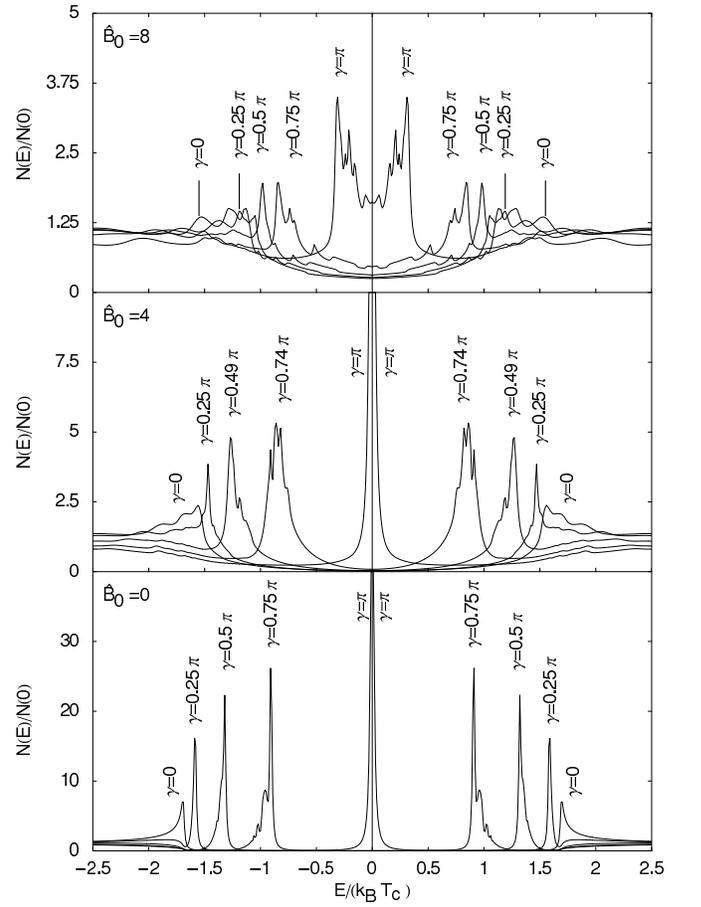}
\end{center}
\caption{\label{cap:ldosB0}Local density of states in the center of the constriction for three different values of the external magnetic field ($\hat{B}_0=0$, $4$ and $8$ from bottom to top). For this figure, parameters are set corresponding to Fig.\,\ref{cap:cphirgapB0}. Please note the different scaling of the vertical axis for the three panels.}
\end{figure}

\begin{figure}
\begin{center}
\includegraphics[width=0.85\columnwidth, keepaspectratio]{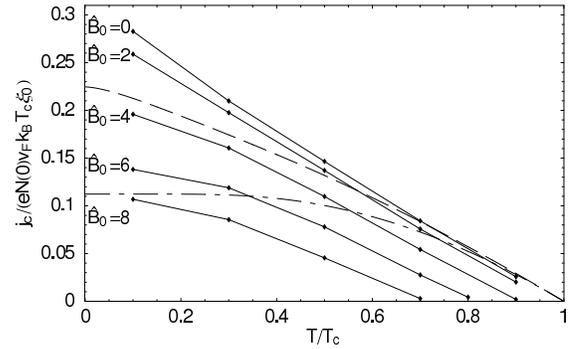}
\end{center}
\caption{\label{cap:jcBT} Temperature dependence of the critical current without magnetic fields ($\kappa=\infty$, $\hat{B}_0=0$, solid guideline) and with external magnetic field ($\kappa=1$, $\hat{B}_0$ as indicated) for $l=w=0.05\,\xi_0$. For comparison, we plot the Kulik-Omel'yanchuk result (\cite{Kul02}, dashed line) and the Ambegaokar-Baratoff result (\cite{Amb01}, dash-dotted).}
\end{figure}

Since our two-dimensional geometry implies translational invariance in the $z$ direction, the high critical field found in our calculations can be understood in terms of the parallel critical field $H_{c3}$~\cite{Sai01,Sco01}. Because of the infinite extension of the geometry in the $z$ direction, the external magnetic field is oriented parallel to the surfaces. Up to an external magnetic field $H_{c3}=1.695\,H_{c2}$ oriented parallel to the surface of a superconductor, there is a superconducting surface sheath of thickness $\sim\xi(T)$ while $\Delta\rightarrow 0$ in the interior. If the geometry consists of two coplanar surfaces with spacing $d<\xi$, nucleation is possible at even higher values of the external magnetic field, growing with $\sim 1/d$ (see \cite{Gel01,Fin01} and references therein). Since the channel width is $W=1\,\xi_0$, the value of the external magnetic field at which superconductivity is suppressed is in good agreement with $H_{c3}$. In the center of the constriction, the spacing between opposite surfaces is even smaller and nucleation occurs at even higher values of the external magnetic field.

From the inset in Fig.\,\ref{cap:cphirgapB0} (b), we conclude that the current across the constriction should be suppressed at about $\hat{B}_0\approx 8.5$ since this value of the external magnetic field leads to suppression of superconductivity in the channel. The section used for the calculations with a constriction ($L=2\,\xi_0$) is too short to reach the undisturbed channel, far enough from the weak link. However, using a longer section would lead to a badly defined phase difference across the junction and to excessive numerical costs. Our main conclusion, the fact that the external magnetic field primarily affects the channel and not the constriction, remains unchanged, however.
\\
In a thin film with small extension in the $z$ direction, the relevant critical field for the suppression of superconductivity is the upper critical field $H_{c2}$. The occurrence of the parallel critical field $H_{c3}$ in our results stems from the assumption of translational invariance in the $z$ direction. 

In Fig.\,\ref{cap:ldosB0}, we plot the local density of states in the center of the constriction for three different values of the magnetic field. At intermediate strength of the external magnetic field, we find a splitting of the Andreev bound states, but the main features of the LDOS without a magnetic field persist. We still find the energy gap and pronounced bound states if a phase difference $\gamma\neq 0$ is applied. For $\gamma=\pi$, the bound states have zero energy which indicates Josephson junction behaviour. For very strong external magnetic fields, the local density of states is strongly disturbed. Even without a phase difference, the energy gap is partly filled. A phase difference still leads to bound states, but their spectral weight is strongly reduced. For $\gamma=\pi$, the bound states have finite energy and the Josephson effect is absent.

In Fig.\,\ref{cap:jcBT}, we show the temperature dependence of the critical current for $l=w=0.05\,\xi_0$ without ($\kappa=\infty$, $\hat{B}_0=0$) and with external magnetic field ($\kappa=1$, $\hat{B}_0$ as indicated in the figure). For comparison, we plot the Kulik-Omel'yanchuk result for a point contact in the clean limit~\cite{Kul02} and the Amgebaokar-Baratoff result for a tunnel junction~\cite{Amb01}. We find that the universality of the inclination of $j_c(T)$ close to $T_c$ persists even in an external magnetic field.

\section{\label{sec:concl}Conclusions}

We present self-consistent solutions of microscopic Eilenberger theory for a two-dimensional model of a ScS Josephson junction. Magnetic fields, external ones as well as those evoked by the supercurrents, have been included and the relevant equations have been solved numerically without further assumptions.

In the self-consistent calculation, the Josephson effect appears without further input. Unlike in a non-self-consistent calculation, we do not have to make assumptions about the existence and the behaviour of the weak link. By considering the self-consistent results, we explain in detail how the Josephson effect in a ScS junction emerges. Taking into account the order parameter amplitude and phase, the currents and the local density of states, we point out that the Josephson effect is a result of phase-coherent quasiparticle transport across the junction. Localized quasiparticle states (Andreev bound states) appear in the junction if a supercurrent flows. These bound states, together with the corresponding current-phase relations, have been calculated and discussed in detail for different geometric parameters, different temperatures, different values of $\kappa=\lambda_L/\xi_0$ and different values of the external magnetic field.

We show that the Josephson effect in ScS junctions is independent of $\kappa$ as long as the London penetration depth is larger than or comparable to the size of the constriction. We also show that the Josephson effect in ScS junctions is very insensitive to external magnetic fields. External magnetic fields mainly influence the channel leading to the weak link and suppress superconductivity there. The Josephson effect persists up to the upper critical field of the channel.

In contrast to the non-self-consistent results, the self-consistently calculated local density of states exhibits Andreev bound states with peculiar substructure due to the geometry and magnetic fields. The local density of states should be experimentally accessible via STM/STS.

\acknowledgments

We would like to thank S.~Graser, C.~Iniotakis and E.~H.~Brandt for valuable discussions.


\begin{thebibliography}{10}
\bibitem{foot1}In the present work, we use $\delta=0.01\,k_B T_c$ throughout.
\end{thebibliography}
\end{document}